\documentclass[traditabstract]{aa}
\usepackage{graphicx}
\usepackage{lscape}
\usepackage{placeins}
\usepackage{txfonts}
\usepackage{ulem}
\usepackage{natbib}
\usepackage{pifont}
\usepackage{xcolor}
\usepackage{amsmath}
\usepackage{dsfont}

\usepackage{xspace}

\newcommand{\integral}{\textit{INTEGRAL}\xspace}

\newcommand{\maxi}{\textit{MAXI}\xspace}

\newcommand{\kTz}{\ensuremath{k\mathrm{T}_\mathrm{0}}\xspace}
\newcommand{\kT}{\ensuremath{k\mathrm{T}}\xspace}

\newcommand{\refl}{\ensuremath{\Omega/2\pi}\xspace}
\newcommand{\EFe}{\ensuremath{E_\mathrm{Fe}}\xspace}
\newcommand{\sigmaa}{\ensuremath{\sigma_\mathrm{Fe}}\xspace}

\newcommand{\chired}{\ensuremath{\chi^2_\mathrm{red}}\xspace}
\newcommand{\Gammapo}{\ensuremath{\Gamma_\mathrm{po}}\xspace}
\newcommand{\Gammath}{\ensuremath{\Gamma_\mathrm{th}}\xspace}
\newcommand{\nh}{\ensuremath{n_\mathrm{H}}\xspace}

\newcommand{\quinze}{MAXI J1535$-$571\xspace}
\newcommand{\dixhuit}{MAXI J1820+070\xspace}
\newcommand{\treize}{MAXI J1348$-$630\xspace}

\makeatletter
\newcommand{\srcsize}{\@setfontsize{\srcsize}{7pt}{8pt}}
\makeatother

\begin{document} 

   \title{\integral study of \quinze, \dixhuit and \treize outbursts}
   \subtitle{I. Detection and polarization properties of the high-energy emission}

   \author{F. Cangemi\inst{1,2} \and
          J. Rodriguez\inst{3} \and
          T. Belloni\inst{4} \and
          C. Gouiffès\inst{3} \and
          V. Grinberg\inst{5} \and
          P. Laurent\inst{3} \and
          P.-O. Petrucci\inst{6} \and
          J. Wilms\inst{7}
          }

\offprints{cangemi@apc.in2p3.fr}
\authorrunning{Cangemi et al.}
\titlerunning{High-energy properties of \quinze, \dixhuit and \treize}

   \institute{Sorbonne Université, CNRS/IN2P3, Laboratoire de Physique Nucléaire et de Hautes Energies, LPNHE, 4 place Jussieu, 75005 Paris, France\\ \email{cangemi@apc.in2p3.fr}
   			\and
   			Université Paris-Cité, AstroPrticules et Cosmologie, APC, 10 rue Alice Domon et Léonie Duquet, 75013 Paris, France
             \and 
             Université Paris-Saclay, Université Paris-Cité, CEA, CNRS, AIM, 91191, Gif-surYvette, France
             \and
              INAF-Osservatorio Astronomico di Brera, via E. Bianchi 46, I-23807, Merate, Italy       
             \and 
             European Space Agency (ESA), European Space Research and Technology Centre (ESTEC), Keplerlaan 1, 2201 AZ Noordwijk, the Netherlands
             \and
             Institut de Planétologie et d'Astrophysique de Grenoble, Université de Grenoble Alpes, 38000 Grenoble, France
             \and
             Dr. Karl Remeis-Sternwarte and Erlangen Centre for Astroparticle Physics, Friedrich-Alexander Universität Erlangen-Nürnberg, Sternwartstr. 7, 96049 Bamberg, Germany
         }
   \date{Accepted;}

  \abstract{In black hole X-ray binaries, a non-thermal high-energy component is sometimes detected at energies above 200\,keV. The origin of this high-energy component is debated and distinct spectral modelizations can lead to different interpretations. High-energy polarimetry measurements with \integral allow new diagnostics on the physics responsible for the MeV spectral component in black hole X-ray binaries.}
  {In this work, we aim to investigate the high-energy behavior of three bright sources discovered by the Monitor of All-sky X-ray Image: \quinze, \dixhuit and \treize. We take advantage of their brightness to investigate their soft $\gamma$-ray (0.1--2\,MeV) properties with \integral. We use both spectral and polarimetric approaches to probe their high-energy emission with the aim to bring new constraints on the $\sim$ MeV emission in black hole X-ray binaries.}
  {We first study the spectral characteristics of the sources in the 3--2000\,keV using JEM-X, IBIS and SPI with a semi-phenomenological description of the data. We then use IBIS as a Compton telescope in order to evaluate the polarization properties of the sources above 300\,keV.}
  {A high-energy component is detected during the Hard-InterMediate State and Soft-Intermediate State of \quinze, the Low-Hard State of \dixhuit and the Low-Hard State of \treize. The components detected in \dixhuit and \treize are polarized with a polarization fraction of $26 \pm 9$\textdegree\ and $> 56$\,\% in the 300--1000\,keV, respectively. With no polarization information for \quinze, the component detected could either come from the jets rather than the corona. In the case of \dixhuit, the extrapolation of the synchrotron spectrum measured in the infrared indicates that the component is likely due to a non-thermal distribution of electrons from a hybrid corona. For \treize, the high fraction of polarization points towards a jets origin, however, we cannot formally conclude without any infrared data giving information on the optically thin part of the synchrotron spectrum.}{}

   \keywords{Accretion, accretion disks --- Physical data and processes --- Black hole physics --- X-rays: binaries --- Stars: black holes}

   \maketitle
%

\section{Introduction}

Black hole X-ray binaries are transient systems that can transit through various spectral states during their outburst. The two main states are denoted the Low Hard State (LHS) and the High Soft State \citep[HSS; see][for a precise definition of spectral states]{Remillard2006, Belloni2010}. The LHS corresponds to the rising phase of the outburst; its 1--200\,keV spectrum can be well described by a powerlaw with a photon index of $\Gamma \sim$ 1.5 with an exponential cutoff usually around 100\,keV. This component is usually interpreted as the emission from Compton scattering of disk photons by electrons from a hot plasma called the \textquotedblleft corona\textquotedblright. These sources are also sites of ejections of material at relativistic speeds, and sources of strong disk winds. The discovery of jets \citep{Mirabel1992} led to name them microquasars. A small extension, steady \textquotedblleft compact\textquotedblright\ jet is observed in the LHS, and has been resolved in a few sources: Cygnus X--1, GRS 1915+105, and more recently, MAXI J1348--630 \citep[e.g.,][]{Stirling2001, Fuchs2003, Carotenuto2021}. Transitions from the harder to the softer states are accompanied by transient, discrete, and large scale ejections \citep[e.g.,][]{Fender1999, Hannikainen1999, Mirabel1998, Corbel2001, Rodriguez2008a} and no jet seems to persist in the HSS (although see \cite{Rushton2012, Zdziarski2020} for the case of Cygnus X-1). The X-ray flux in the HSS is dominated by the photon disk emission peaking at $\sim$ 1\,keV. This component is associated with a continuum usually described with a powerlaw with a photon index $\Gamma > 2.5$ and whose origin is still not yet understood.

In several sources, observations made at higher energies ($> 200$\,keV) have revealed the presence of a high-energy non-thermal component extending up to 1\,MeV \citep{Grove1998, CadolleBel2006, Laurent2011, Rodriguez2015, Cangemi2021b}. The origin of this high-energy emission is still not well understood and (at least) two scenarios have been invoked to explain its origin. In the first scenario, this component is the  extension of the Synchrotron spectrum from the basis of the jets \citep{Markoff2005, Laurent2011, Jourdain2014, Rodriguez2015, Kantzas2021}. The alternative explanation suggests that this component arises from a non-thermal distribution of electrons in a hybrid thermal/non-thermal corona \citep[e.g.,][]{DelSanto2013, Romero2014, Cangemi2021a, Cangemi2021b}. In the case of the high mass BHB, Cygnus X-1, \cite{Cangemi2021b} have suggested that the high-energy tails, seen both in the HSS and LHS have different origins: in the HSS an hybrid corona is favoured while the jet would be at the origin of the LHS high-energy tail. 

Polarization measurements of the high-energy emission from BHBs is probably the best way to disentangle between these different scenarios as we expect distinct polarization properties between a Compton or synchrotron emission. \integral has already brought important insights on polarization measurements thanks to the design of the SPectrometer on board on \integral \citep[SPI,][]{Vedrenne2003} and the Imager on Board on \integral Satellite \citep[IBIS;][]{Ubertini2003} which can be used as a Compton telescope with its two layers plane (ISGRI and PICsIT). Indeed the high polarization degree of the $>400$\,keV Cygnus X-1 high-energy tail detected both with IBIS and SPI \citep{Laurent2011, Jourdain2012} has be shown to be compatible with the emission of the compact jet in the LHS of this source \citep{Rodriguez2015}. Cygnus X-1 is currently the only source for which polarization measurements have been able to constrain the origin of the high-energy component. It is therefore important to look at other sources in order to probe the origin of this component and try to obtain a more general understanding of the $\sim$ MeV emission in microquasars. 


\quinze, \dixhuit and \treize are X-ray transients discovered by the Monitor of All-Sky X-ray Image (\maxi) on board on the International Space Station \citep{Matsuoka2009} during their outburst in September 2017 \citep{Negoro2017}, March 2018 \citep{Kawamuro2018} and January 2019 \citep{Yatabe2019} respectively. The three outbursts were then followed by \integral \citep[e.g.,][]{Lepingwell2018, Bozzo2018, Cangemi2019a, Cangemi2019b} and all three sources were particularly bright and reached a maximum flux of a few Crabs. Not many other sources have reached such a brightness previously, and therefore, polarization studies on short accumulation (i.e. a few days) of data led to totally unconstrained results. Even for Cygnus X-1 the detection of polarised emission required accumulation of Ms of data over large intervals of time.




Although the mass of the compact object in \quinze has not been established yet, bright radio emission associated with a flat radio spectrum \citep{Russell2017} and strong emission in the infrared band \citep{Dincer2017, Vincentelli2021} led to identify the source as a BHB. The source displays strong X-ray variability, including low-frequency Quasi-Periodic Oscillations \citep[QPOs, ][]{Stiele2018, Huang2018, Stevens2018, Bhargava2019, Sreehari2019}. \dixhuit is a X-ray binary harbouring a black hole of mass $\sim 8.5$\,M$_\odot$ accreting from a companion star of $\sim 0.4$\,M$_\odot$ \citep{Torres2019,Torres2020}. The intense brightness of the system has triggered multiple multi-wavelength observing campaigns \citep[e.g.,][]{Bright2020, Trushkin2018, Tetarenko2021, Hoang2019} and the source was the center of many studies. 
\treize is also a BHB with a black hole of mass $\sim 11$\,M$_\odot$ \citep{Lamer2020} and the source is located at $\sim 3.3$\,kpc according to the measurement of \cite{Lamer2020} with \textit{SRG/eROSITA} and \textit{XMM-Newton} observations. The source shows also strong X-ray variability and QPOs were observed during LHS to HSS transition \citep{Belloni2020}. The source shows also a interesting behavior when looking at is radio/X-ray correlation which infers the relation between the emission from the compact jets and the inner accretion flow \citep{Corbel2013}. The correlation displays two different tracks reffered as the \textit{standard} and the \textit{outliers} tracks. Usually, a source follows one of the tracks during its outburst, but \cite{Carotenuto2021} have shown that in the the case of \treize, the source follows the outliers track during the first part of the outburst before joining the standard track. Table \ref{tab:source_characteristics} summarises all the known parameters for the three sources.


\begin{table*}

\centering
\caption{Summary of the known parameters for the three studied sources.}
\begin{tabular}{l c c c}
\hline \hline
Source & \quinze & \dixhuit & \treize \\ [1.5pt]
\hline
Mass of the black hole [M$_\odot$] & -- & $8.5^{+0.8}_{-0.7}$  & $11^{+2}_{-2}$  \\ [1.5pt]
 & -- & \cite{Torres2020} & \cite{Lamer2020} \\ [1.5pt]
Mass of the companion [M$_\odot$] & -- & $0.6^{+0.1}_{-0.1}$ & -- \\ [1.5pt]
& -- & \cite{Torres2020} & -- \\ [1.5pt]
Distance [kpc] & $4.1^{+0.6}_{-0.5}$  & $3.0^{+0.3}_{-0.3}$ & $3.3^{+0.3}_{-0.3}$\\ [1.5pt]
& \cite{Chauhan2019} & \cite{Atri2020} & \cite{Lamer2020} \\ [1.5pt]
Inclination [°]& $67.4^{+0.8}_{-0.8}$ & $63^{+3}_{-3}$ & $28^{+3}_{-3}$ \\ [1.5pt]
& \cite{Miller2018} & \cite{Torres2020} & \cite{Anczarski2020} \\[1.5pt]

\hline

\label{tab:source_characteristics}
\end{tabular}
\end{table*}




In this study, we present the evolution of the outbursts of those three different sources as observed by \integral. We make use of its unique capabilities in order to probe the main properties over the full 3–1000 keV range covered by the observatory. We also present polarization measurements of the high-energy component when significant emission is detected above 300\,keV. The description of the observations, and the data reduction methods are reported in Sect. \ref{sec:obs_red}. Section \ref{sec:spectral} is dedicated to the phenomenological spectral analysis. We then present the results from our polarization measurements with the Compton mode above 300\,keV for the three sources in Sect. \ref{sec:polarization}. The results are finally discussed in Sect \ref{sec:interpretation}.

\section{Data reduction}
\label{sec:obs_red}

\subsection{Spectral extraction: \integral/JEM-X}
We only use data from JEM-X unit 1. The data are reduced with the \textsc{Off-Line Scientific Analysis} (\textsc{osa}) version 11.1 software. We follow the standard steps described in the JEM-X user manual\footnote{\url{https://www.isdc.unige.ch/integral/download/osa/doc/11.1/osa_um_jemx/man.html}.}. Spectra are extracted for each SCience Windows\footnote{\integral individual pointings of a duration of 1800--3600 s typical duration.} (scw heareafter) where the source is automatically detected by the software at the image creation and fitting step. Spectra are then computed over 32 logarithmic spectral channels from $\sim 3$ to $\sim 34$\,keV using the standard binning definition. Individual spectra for each period are then combined with the \textsc{OSA} \textit{spe\_pick} tool according to the classification scheme described in Sect. \ref{sec:classification}. The appropiate ancillary response files (arfs) are produced during the spectral extraction and are combined with \textit{spe\_pick}, while the redistribution matrix file (rmf) is rebinned from the instrument characteristic standard rmf with \textit{j\_rebin\_rmf}. We add 3\,\% systematic error on all spectral channels for each of the stacked spectra, as recommended in the JEM-X user manual. For our spectral analysis, we consider the spectra from 3 to 20\,keV. 


\subsection{Spectral extraction: \integral/IBIS/ISGRI}
Data from \integral/IBIS upper detector ISGRI are also reduced with the \textsc{osa 11.1} software. We use the standard procedure described in the IBIS manual\footnote{\url{https://www.isdc.unige.ch/integral/download/osa/doc/11.1/osa_um_ibis/man.html}.}. For each scw, we create the sky model and reconstruct the sky image and the source count rates by deconvolving the shadowgrams projected onto the detector plane. For the three sources, spectra are extracted using 60 logarimically spaced channels between 20\,keV and 1000\,keV. Response matrixes are automatically generated running the \textsc{osa 11.1} spectral extraction. We then use the \textit{spe\_pick} tool to create stacked spectra for each different outburst periods (see Sect. \ref{sec:classification}). 2\,\% of systematics are added to each stacked spectra as indicated in the IBIS user manual.

\subsection{Spectral extraction \integral/SPI}

We followed a similar procedure already applied in the case of Cygnus X-3 \citep{Cangemi2021a}. To summarize, we use the \textsc{Spi Data Analysis Interface}\footnote{\url{https://sigma-2.cesr.fr/integral/spidai}.} (\textsc{Spidai}) to reduce SPI data. During the outburst of \treize, SPI was in annealing from MJD 58501 to MJD 58522 (\integral revolutions 2047 to 2054), therefore we do not have available SPI data during the LHS and the IMS but we extract the SPI spectrum in the HSS. We respectively create a sky model containing \quinze, \dixhuit and \treize and set their variability to 5 scws. We then create the background model by setting the variability timescale of the normalization of the background pattern to 5 scws. Sometimes, solar flares, radiation belt entries, and other non-thermal incidents can lead to unreliable results. Therefore, we remove scws for which the reconstructed count compared to the detector counts give a poor $\chi^2$ (\chired > 1.5) in order to avoid these effects. This selection reduces the total number of scws by $\sim 10$\,\%. The shadowgrams are then deconvolved to obtain the source flux, and spectra are extracted between 20\,keV and 2000\,keV using 27 logarithmically spaced channels. For each spectrum, we apply a correcting factor of 1/0.85 above 400\,keV in order to take into account a change efficiency above this threshold \citep{Roques&Jourdain2019}.
 
\subsection{Data selection and period definition}
\label{sec:classification}

\begin{figure*}[t]
    \centering
    \includegraphics[width=\textwidth]{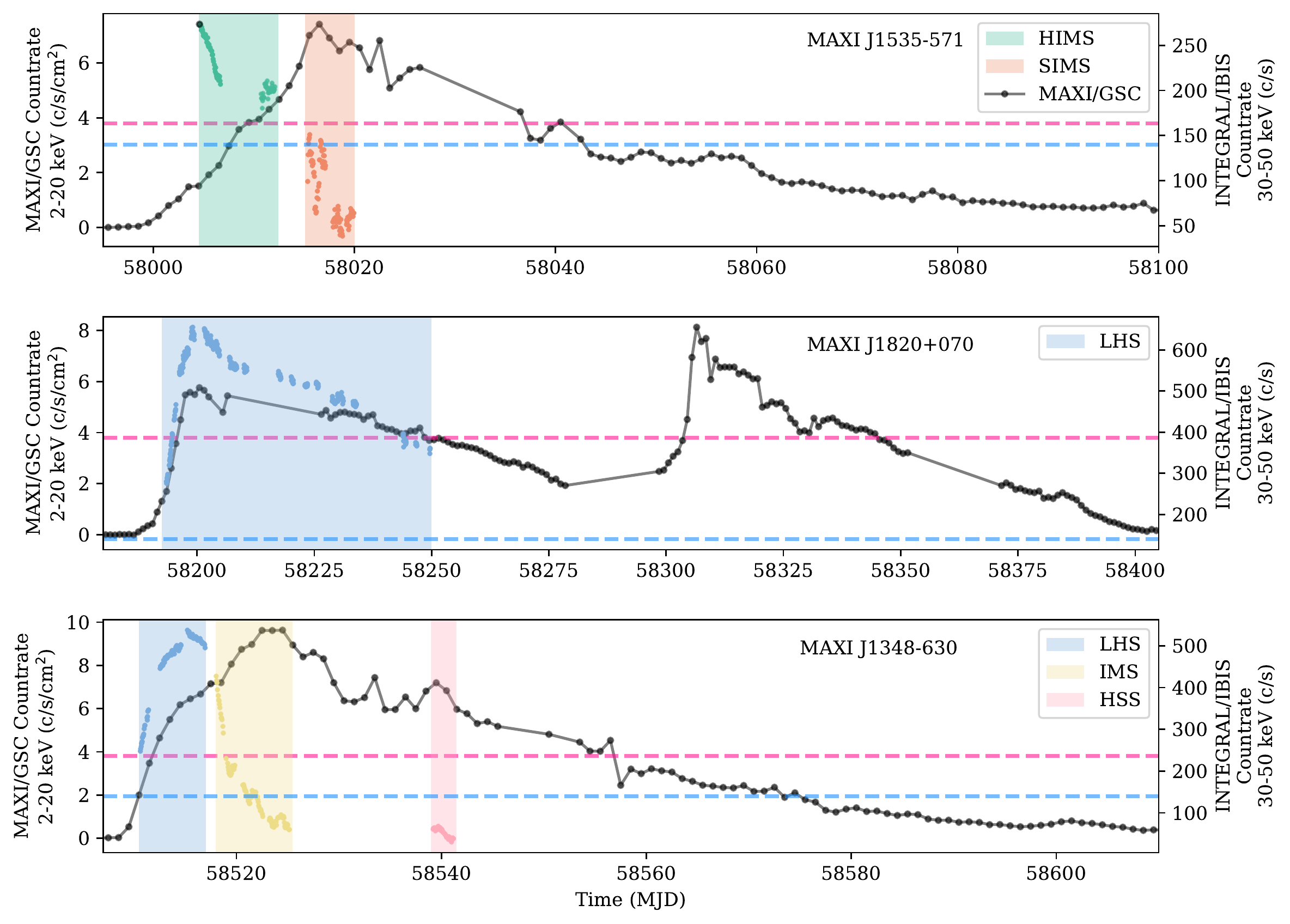}
    \caption{\maxi/GSC 2--20\,keV (in black) and \integral/IBIS 30--50\,keV (colored dots) lightcurves of \quinze (top panel), \dixhuit (middle panel), and \treize (bottom panel). Left and right y axis indicate the countrate for \maxi/GSC and \integral/IBIS respectively. Definitions of the different periods and the corresponding colors are described in Sect. \ref{sec:classification}. The counrates for 1 Crab is indicated with the pink and blue dotted lines for \maxi/GSC and \integral/IBIS, respectively.}
    \label{fig:lightcurves}
\end{figure*}

We consider scw between MJD 58004--58019 (09/08/2017--09/23/2017), 58193--58249 (03/16/2018--05/11/2018) and 58512--58541 (01/29/2019--02/27/2019) respectively corresponding to the outburst of \quinze,  \dixhuit, and \treize where the sources are less than 10\textdegree\ off axis. 
Figure \ref{fig:lightcurves} shows the \integral/ISGRI lightcurves between 30\,keV and 50\,keV along with the \maxi/GSC monitoring in order to show the \integral observations in the long-term context of the whole outburst.

For the three sources, we perform purely phenomenological fit for each scw, in the 3--300\,keV energy range using JEM-X unit 1 and ISGRI in order to separate the three datasets into different periods according to their spectral shapes and properties. The fits are performed using \textsc{Xspec} \citep{Arnaud1996}; we use a cutoff powerlaw model (\textsc{cutoffpl}) and added the emission from a disk (\textsc{diskbb}) when needed.
 We also multiply the model with a constant (\textsc{constant}) allowing us to take into account calibration issues between the instruments and the differences in total exposure between the instruments\footnote{Given the \integral observing pattern around the pointed source and since IBIS has a larger field of view than JEM-X, sources can be outside the JEM-X field of view while still in the IBIS one)}. Constants are let free whereas other parameters are tied between instruments.
 
\begin{figure*}[h!]
	\centering
	\includegraphics[width=\textwidth]{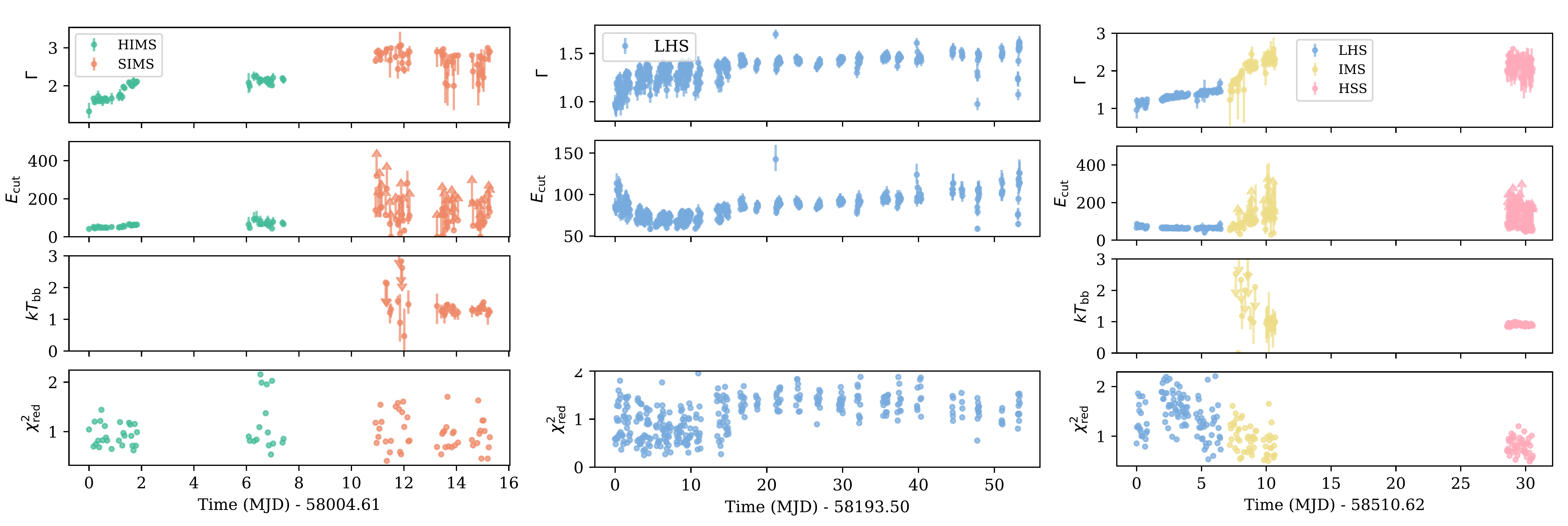}
	\caption[width=\textwidth]{Temporal evolution of the different spectral parameters extracted from our phenomenological spectral fitting for \quinze (left panel), \dixhuit (middle panel), and \treize (right panel). Definitions of the different periods and their corresponding colors are described in Sect. \ref{sec:classification}}\label{fig:evol}	
\end{figure*}

In summary, we use \textsc{constant*cutoffpl} or \textsc{constant*(cutoffpl+diskbb)} when high residuals are observed at low energy. Because of the low exposure of the individual spectra, we do not use absorption here, the aim being to roughly describe the shape of the spectra in order to separate the different periods of the outbursts. We then extract the values of the photon index, the exponential cutoff and the disk temperature. The evolution of these different parameters are shown in Fig. \ref{fig:evol} for the three sources.

For \quinze, based on the value of the photon index which transits from $\sim 1.9$ to $\sim 2.3$ from MJD 58004 to MJD 58016, we divide our datasets in two periods: 
the first one from 58004 to 58012 and the second one, after MJD 58016.
 The two periods are shown in green and orange on Fig. \ref{fig:lightcurves}. This subdivision is consistent with the epochs defined in \citet{Russell2020} where they find that the source was in a Hard InterMediate State (HIMS) state from MJD 58008 to 58016 before its transition into a Soft Intermediate State (SIMS) at MJD 58017. Therefore, we name our two periods according to this classification.

Concerning \dixhuit, the values of the photon index and the exponential cutoff are consistent with typical values observed in a LHS during all the \integral observations. We thus define a unique period for the characterization of this outburst. The unique period is shown in blue in Fig. \ref{fig:lightcurves}. This is consistent with the different epochs defined in \cite{Buisson2019} where they show that the source starts its transition into the HSS around MJD 58306, after our observations.

The study of the \treize spectral evolution shows two changes in the spectral shape. The first change occurs around MJD 58517 where we observe an increase of the photon index from $\sim$1.6 to 2.3 and an increase of the cutoff energy from 50\,keV to above detection threshold. We define these two periods as the LHS (in blue of Fig. \ref{fig:lightcurves}) and InterMediate State (IMS, in yellow on Fig. \ref{fig:lightcurves}) states. Here we are not able to identify the different flavors (HIMS and SIMS) of the IMS. The observations from the last period, after MJD 58539, show a higher photon index value $\sim 2.3$ associated with a high-energy exponential cutoff, not constrained in the majority of our observations (see the lower limits on Fig \ref{fig:evol}). 
\citet{Belloni2020} find a similar state classification when analysing data from \textit{NICER}. Figure \ref{fig:lightcurves} shows the LHS, IMS and HSS state periods in blue, yellow and pink, respectively. 



\begin{figure*}[t]
    \centering
    \includegraphics[width=\textwidth]{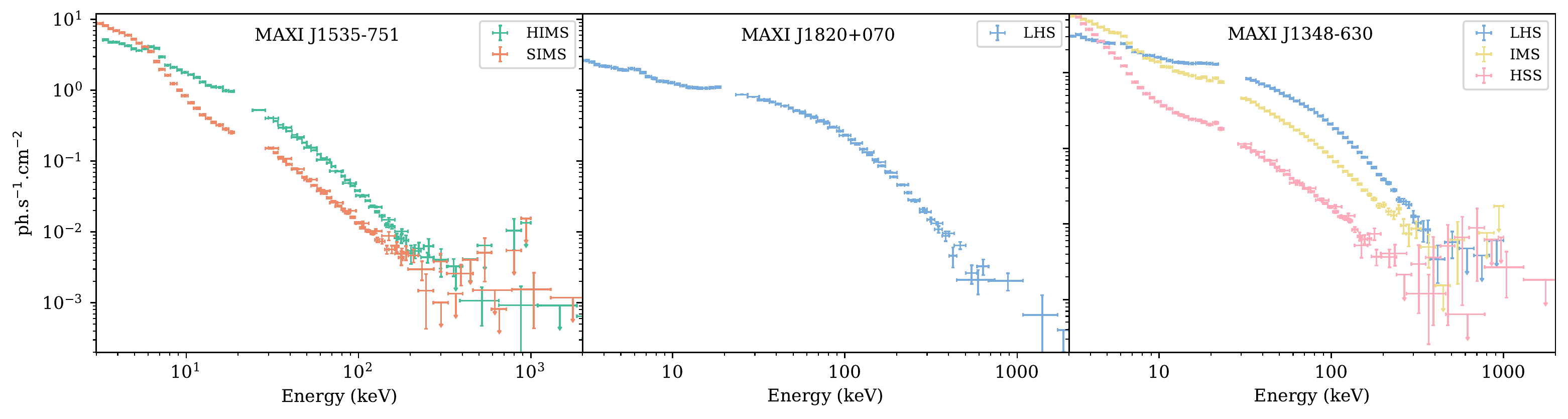}
    \caption{Stacked spectra extracted from JEM-X, ISGRI and SPI for \quinze (left panel), \dixhuit (middle panel), and \treize (right panel). The different color indicates the different epochs considered using the same color code as in Fig. \ref{fig:lightcurves}.}
    \label{fig:spectra}
\end{figure*}

The broadband \integral spectra are shown on Fig. \ref{fig:spectra}. The different state spectra are plotted with different colors using the same color code as in Fig. \ref{fig:lightcurves}.

\section{State dependent spectral analysis}
\label{sec:spectral}

\subsection{Methodology}
 
 According to the state classification described in Sect. \ref{sec:classification}, we have different stacked spectra for different periods of the three outbursts. In order to assess the potential presence of a high-energy tail in a state-resolved way, we follow the same methodology described in \cite{Cangemi2021a, Cangemi2021b}. This methodology is divided in two steps: we first analyse the data from 3 to 100 keV with a reflected \citep[\textsc{reflect}, ][]{Magdziarz1995} thermal Comptonization continuum \citep[\textsc{nthcomp}, ][]{Zdziarski1996}. We also add the iron line at 6.4\,keV with a Gaussian (\textsc{gaussian}). Although, the reflection is not needed in the individual spectra, it is needed in the stacked spectra in order to obtain a statistically good fit. We choose this Comptonization model rather than another because the \dixhuit spectrum is better fitted with this model than with other thermal Comptonization models we have tested (e.g., \textsc{comptt}). This model has also been used by \cite{Shidatsu2019} to fit \textit{MAXI}/GSC and \textit{Swift}/BAT data of \dixhuit. For the sake of consistency, we also use \textsc{nthcomp} for \quinze and \treize. We also add the absorption by the interstellar medium \citep[\textsc{tbabs}, ][]{Wilms2000} using \textsc{angr} solar abundances \citep{AndersEbihara1982}. The model is written \textsc{constant*tbabs*(reflect(nthcomp) + gaussian)} in \textsc{Xspec}. We add the emission from a disk \citep[\textsc{diskbb},][]{Mitsuda1984} when needed. In this case, the model becomes \textsc{constant*tbabs*(reflect(nthcomp) + diskbb + gaussian}). We let the energy of the iron line vary between 6.2\,keV to 6.5\,keV whereas its width is allowed to vary between 0.2\,keV to 0.5\,keV. Once we obtain a satisfactory fit, we add the data above 100\,keV, let the parameters vary freely and search for presence of residuals at high-energy. In case of large residuals observed above 300\,keV, we add a powerlaw component (\textsc{constant*tbabs*(reflect(nthcomp) + gaussian + powerlaw)} or \textsc{constant*tbabs*(reflect(nthcomp) + diskbb + gaussian + powerlaw}) to the model and investigate the significance of this component by performing a F-test. We use an inclination of $i = 67$° \citep{Miller2018}, 63° \citep{Torres2019}, and 28° \citep{Anczarski2020} for the reflection component for \quinze, \dixhuit and \treize respectively. The reflection fraction is allowed to vary between 0 < \refl < 2.

\begin{figure*}[t]
    \centering
    \includegraphics[width=0.6\textwidth]{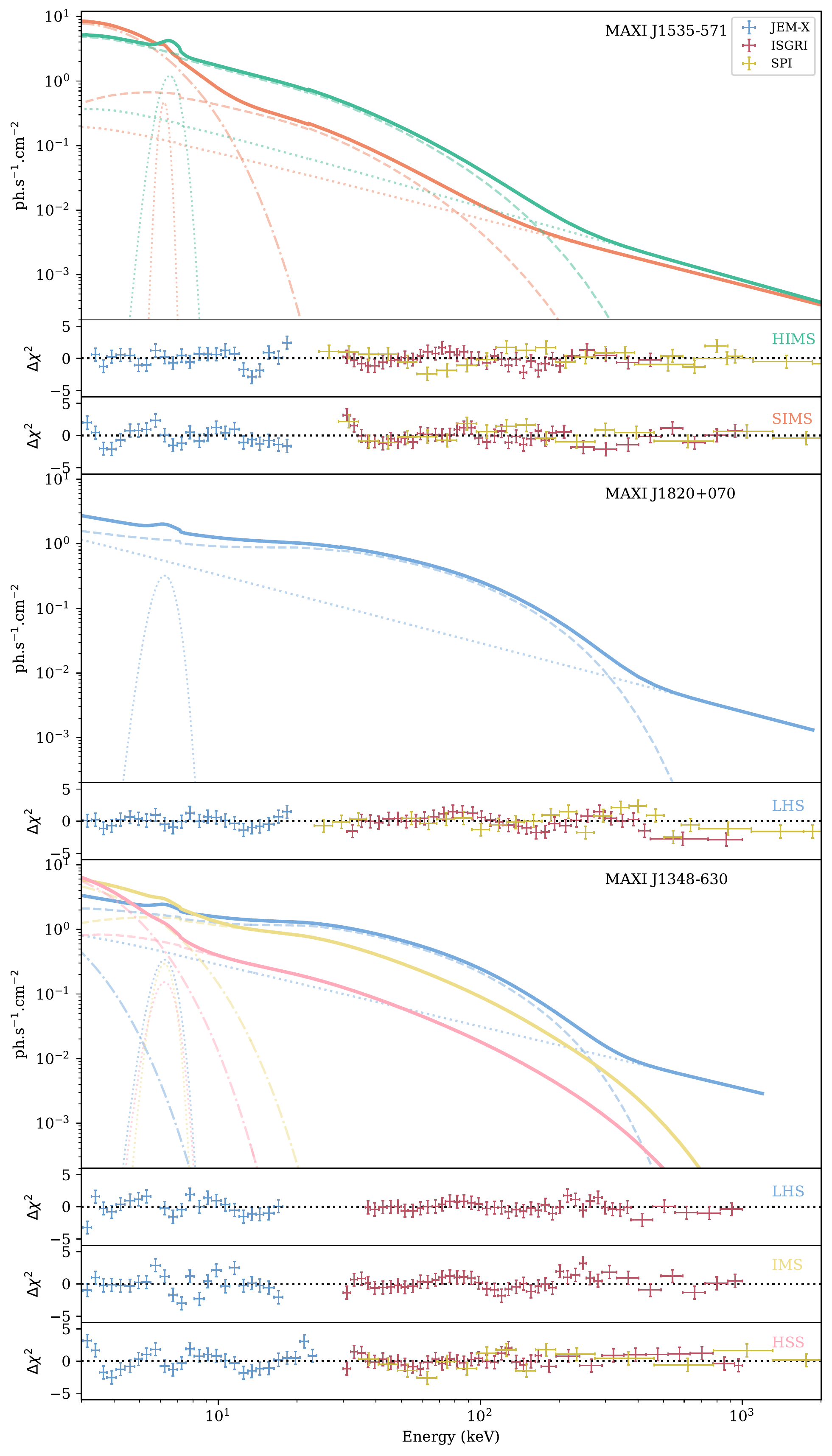}
    \caption{Best-fit models obtained from our phenomenological fitting of the 3--2000\,keV for \quinze (top panel), \dixhuit (middle panel), and \treize (bottom panel). Different colors indicate different epochs using the same color code as in Fig. \ref{fig:lightcurves}. JEM-X, ISGRI and SPI data are represented with blue, red and yellow respectively. Residuals for each period are plotted separately at the bottom of the corresponding spectrum. Different model components are shown with different line styles: Comptonized continnum,(dashed), disk (dotted-dashed), additional powerlaw (densely dotted), gaussian (dotted).}
    \label{fig:spectra_fits}
\end{figure*}

The best-fit parameters obtained in the 3--2000\,keV band are reported in Tab. \ref{tab:spectral_analysis} and the corresponding spectral fits for each source in the different states are shown in Fig. \ref{fig:spectra_fits}. If the addition of a disk is necessary to fit the data, we indicate \textquotedblleft yes\textquotedblright\xspace in the row called \textquotedblleft Disk\textquotedblright\xspace. Rows \textquotedblleft Flux\textquotedblright\xspace and \textquotedblleft Flux$_\mathrm{po}$\textquotedblright\xspace respectively refer to the flux given by the total model and the flux which comes from an additional powerlaw component in the 300--1000\,keV range.


\begin{table*}[t]
    \caption{Parameters obtained for our phenomenological fitting for each source and each period.}
    \label{tab:spectral_analysis}
    \centering
    \begin{tabular}{l c c c c c c}
    \hline \hline
        Parameters & \multicolumn{2}{c}{\quinze} & \dixhuit & \multicolumn{3}{c}{\treize} \\
        \hline
		 & HIMS & SIMS & LHS & LHS & IMS & HSS \\
		 \hline
		 & & & & & & \\
		 C$_\mathrm{ISGRI}$ & $0.93^{+0.04}_{-0.03}$ & $1.14^{+0.01}_{-0.02}$ & $0.91^{+0.03}_{-0.02}$ & $0.82^{+0.03}_{-0.02}$ & $0.8^{+0.2}_{-0.2}$ & $0.85^{+0.04}_{-0.04}$\\ [1.5pt]
		 C$_\mathrm{SPI}$ & $0.90^{+0.04}_{-0.03}$ & $1.20^{+0.01}_{-0.02}$ & $0.91^{+0.03}_{-0.02}$ & -- & -- & $0.90^{+0.04}_{-0.04}$\\  [1.5pt]
		 \nh [$\times 10^{22}$ cm$^{-2}$] & $3.3^{+0.6}_{-0.7}$& $2.5^{+1}_{-1}$ & 0.14\,F & 0.86\,F & 0.86\,F & 0.86\,F \\
		\kTz [keV] &  0.3\,F & $1.24^{+0.04}_{-0.03}$ & 0.2\,F &  0.5\,F & $1.21^{+0.09}_{-0.08}$ & $0.84^{+0.03}_{-0.03}$ \\  [1.5pt]
		\kT [keV] & $52^{+43}_{-17}$ & $40^{+106}_{-26}$ & $57^{+4}_{-4}$ &  $44^{+5}_{-4}$& > 193 & > 471 \\  [1.5pt]
		\refl & $1.1^{+0.8}_{-0.3}$ & > 1.8 & > 1 & $0.8^{+0.1}_{-0.1}$ & $1.0^{+0.2}_{-0.1}$ & $0.46^{+0.09}_{-0.08}$ \\ [1.5pt]
		\EFe [keV] & $6.5^{+0.0}_{-0.1}$ & $6.2^{+0.3}_{-0.0}$ & $6.2^{+0.3}_{-0.0}$ & $6.20^{+0.02}_{-0.00}$ & $6.20^{+0.09}_{-0.00}$ & $6.2^{+0.3}_{-0.00}$ \\ [1.5pt]
		\sigmaa [keV] & $0.5^{+0.0}_{-0.2}$ & $0.2^{+0.3}_{-0.0}$ & $0.5^{+0.0}_{-0.3}$ & $0.5^{+0.0}_{-0.2}$ & 0.4\,F & $0.5^{+0.0}_{-0.2}$ \\ [1.5pt]
		\Gammath & $2.38^{+0.05}_{-0.09}$ & $2.52^{+0.08}_{-0.09}$ & $1.61^{+0.01}_{-0.01}$ & $1.69^{+0.04}_{-0.03}$ & $2.06^{+0.03}_{-0.03}$ & $2.19^{+0.02}_{-0.02}$\\ [1.5pt]
		\Gammapo & $2.1^{+0.4}_{-0.5}$ & $2.0^{+0.4}_{-0.4}$ & $2.09^{+0.03}_{-0.03}$ &$2.0^{+0.3}_{-0.3}$ & -- & -- \\ [1.5pt]
		Disk & no & yes & no & yes & yes & yes  \\ [1.5pt]
		Flux 300--1000\,keV [$\times 10^{-9}$\,ergs\,cm$^{-2}$\,s$^{-1}$] & 1.6 & 1.6 & 5.6 & 5.6 & 0.9 & 0.2\\ [1.5pt]
		Flux$_\mathrm{po}$ 300--1000\,keV [$\times 10^{-9}$\,ergs\,cm$^{-2}$\,s$^{-1}$] & 1.6 & 1.6 & 4.6 & 5.3 & -- & -- \\ [1.5pt]
		$\chi^2$/dof & 112.45/91 & 133.88/92 & 88.68/75 & 62.56/68 & 110.18/72 & 135.67/95\\
		 \\
		
        \hline
    \end{tabular}
\end{table*}

\subsection{Results}
\subsubsection{\quinze}


During the HIMS, the seed-photon energy is unconstrained, we thus fix its value to 0.3\,keV according to e.g., \citet{Sridhar2019}, \citet{Tao2018}. The value for the density column \nh $\sim 3.3 \times 10^{22}$\,cm$^{-2}$ is consistent with the value measured by \citet{Sridhar2019} (\textit{Astrosat}) 
but it is slighly lower compared to the value obtained by \citet{Xu2018} (\textit{NuSTAR}). This difference can arise by the use of slighly different epochs of observations between the different analysis. Moreover, \citet{Xu2018} attribute their high value (\nh $\sim 8 \times 10^{22}$\,cm$^{-2}$) to the inclusion of the thermal disk in their modelling. The electron temperature of \kT $\sim 52$\,keV we obtain is consistent with results from \citet{Sridhar2019} and \citet{Tao2018}. We find a photon index of the Comptonized continuum rather soft \Gammath $\sim 2.38$ slighly higher that found in the different modelizations of \citet{Sridhar2019} and a reflection fraction of \refl $\sim 1.1$. During the SIMS, we observe residuals below $\sim$10\,keV and we add a disk to model the corresponding spectra. The disk temperature is tied to the photon seed temperature of \textsc{nthcomp}. We find \kTz $\sim 1.24$\,keV in agreement with the value obtained by \citet{Tao2018}. The other parameters are roughly the same as in the SIMS except for \Gammath $\sim 2.52$ for which we observe a slight increase. Our best fit gives a lower limit of $> 1.8$ for the reflection fraction. We have tested other reflection model like \textsc{relxill} which also give a similarly large reflection fraction. In their study, \cite{Miller2018} have tested different reflection models, and notably \textsc{relxill} and \textsc{relline} and also obtain a large reflection fraction. Following these authors such large value is compatible with an X-ray source very close to the black hole, the large reflection then being due to the strong expected light bending. Our results agree with their best fit parameters. We will not enter in more details of the reflection component here since it is not the main subject of this paper.


Interestingly, we need an addionnal powerlaw component for both states in order to model the spectra above 300\,keV. 
For the HIMS, inclusion of the powerlaw improves \chired from $155.95/93$\,dof to \chired $= 112.45/91$\,dof; for the SIMS, it is improved from \chired $=292.30/94$\,dof to \chired $=133.88/92$\,dof.
 The photon index \Gammapo of this powerlaw is consistent between the two periods, and we find \Gammapo$^\mathrm{HIMS} = 2.1^{+0.4}_{-0.5}$ and \Gammapo$^\mathrm{SIMS} = 2.0^{+0.4}_{-0.4}$. 
We observe that the flux > 300\,keV is dominated by this additional component. In order to assess the existence of this powerlaw component, we also try to fit the data using solely a powerlaw instead of the thermal Comptonization model. In this case, the data is poorly represented (\chired$^\mathrm{HIMS}$ = 490.53/94 dof and \chired$^\mathrm{SIMS}$ = 210.15/95 dof), and we observe strong residuals in the 30--50\,keV range clearly indicating the presence of a cutoff in this energy range. 


\subsubsection{\dixhuit}

In \dixhuit, the value of the density column is unconstrained, thus, we fix it according to the value \nh $=1.4 \times 10^{21}$ obtained by \cite{Kajava2019}.  We find a Comptonized continuum rather hard with a photon index of \Gammath $\sim 1.61$ and an electron energy \kT $\sim 57$\,keV. 
Our value for the electron energy is higher compared to the value found by \citet{Zdziarski2021}, \kT $\sim 12$\,keV, which use the data above 20\,keV from ISGRI and SPI and the 3--80\,keV data from \textit{NuSTAR}. However, they use a modified version of \textsc{compps}\footnote{They use a sinusoïdal distribution of the seed photons in a sphere.} as their thermal description of the continuum and this could explain the difference compared to our study. The value we find for the electron temperature is also roughly consistent with the work of \citet{Chakraborty2020} and \citet{Buisson2019} where they combine \textit{NuSTAR} and \textit{Astrosat} data. They both use a two coronal component model and find \kT $\sim$ 38\,keV when tying the two corona temperature components. We do not add a disk component to the model and we fix the photon seed energy to \kTz = 0.2\,keV \citep[e.g.,][]{Wang2020, Dzielak2021}. We find a reflection fraction \refl > 1.



When adding the data above 300\,keV, the addition of a powerlaw to the model strongly improve the goodness of the fit (from \chired $=367.23/77$\,dof to \chired $=88.68/75$\,dof).
We observe that the > 300\,keV is largely dominated by the emission from this high-energy component (82\,\% of the 300--1000\,keV flux). 



\subsubsection{\treize}

The parameter \nh is unconstrained and we fix its value to \nh $=8.6 \times 10^{21}$cm$^{-2}$ \citep{Tominaga2020}. Parameters found in LHS are very close to those found during the outburst of \dixhuit. We obtain a photon index \Gammath $\sim 1.69$ and an electron energy \kT $\sim 44$\,keV. It is mentioned that a black body component is present in the LHS \citep[e.g.,][]{Tominaga2020, Chakraborty2020, Zhang2021} with a temperature of $\sim 0.5$\,keV. At such a temperature, the flux of the disk could contribute $> 3$\,keV and therefore we add a \textsc{diskbb} component in our model. The normalization of this component is fixed to 12000, value found by \cite{Chakraborty2020}. Strong residuals are observed at high energy when fitting solely with \textsc{const*(reflect(nthcomp) + gauss)}. The addition of a powerlaw component improves \chired from $112.42/70$\,dof to $62.56/68$\,dof.
We find a photon index of \Gammapo = 2.0 and measure a reflection fraction of $\sim$ 0.8.

We also use a disk to model the IMS and the HSS. As for the states of \quinze, the flux at low energy is dominated by the disk emission. The photon disk energy is slighly higher during the IMS than during the HSS. Using \textit{NICER} and \textit{Astrosat}, \citet{Zhang2021} and \citet{Jithesh2021} find a consistent value analysing data from different observations made during the HSS. However, \citet{Zhang2021} observe a higher value of \Gammath ($\sim 3.3$) than observed during both periods corresponding to our HIMS and SIMS. Note that we also model those two periods with a simple reflected powerlaw and a disk (\textsc{reflect(powerlaw) + diskbb + gaussian}). In this case, we do not find a satisfactory fit for the IMS (\chired$^\mathrm{Int}$ = 248.13/75), but the HSS can be well described by this purely phenomenological model and parameters are consistent with the previous modelization.


\section{Polarization with the Compton mode}
\label{sec:polarization}
\subsection{Principle of the Compton Mode}

Thanks to its two layer detectors; ISGRI at the top and the Pixellated Imaging Caesium Iodide Telescope \citep[PICsIT,][]{Labanti2002} at the bottom, IBIS can be used as a Compton polarimeter. 

%
%
%



This concept relies on the cross section $d\sigma$ which represents the probability for a polarized photon with an energy $E_\mathrm{1}$ to enter in interaction with an electron from the detector \citep[e.g.,][]{Evans&Beiser1956}:
\begin{equation}
\label{eq:cross_section}
\frac{d\sigma}{d\Omega} = \frac{r_\mathrm{0}^2}{2}\left(\frac{E_\mathrm{2}}{E_\mathrm{1}}\right)^2 \left(\frac{E_\mathrm{2}}{E_\mathrm{1}} + \frac{E_\mathrm{1}}{E_\mathrm{2}} - 2 \sin^2\theta_\mathrm{c}\cos^2 \phi\right)
\end{equation}
where $E_\mathrm{2}$ is the energy of the scattered photon in the solid angle $d\Omega$, $\theta_\mathrm{c}$ is the scatter angle, $r_\mathrm{0}$ is the electron radius and $\phi$ is the azimuthal angle of the scattered photon with respect to the polarization direction \citep[see e.g., Fig. 1 of][]{Forot2007}. Using the relation between $E_\mathrm{1}$ and $E_\mathrm{2}$:
\begin{equation}
\frac{E_\mathrm{1}}{E_\mathrm{2}} = \frac{1}{1+\frac{E_\mathrm{2}}{m_e c^2} (1 - \cos \theta_\mathrm{c})}
\end{equation}
we note that for a fixed scattered angle, the cross section will be maximal for $\phi = \pi/2 + k\pi$ with $k \in \mathds{Z}$. This creates an asymmetry in the number of detected photons by PICsIT. We can evaluate the detected photon distribution on the PICsIT detector with respect to the azimuth $\phi$:
\begin{equation}
\label{eq:polarigram}
N(\phi) = C[1 + a_\mathrm{0} \cos(2(\phi -\phi_\mathrm{0}))]
\end{equation}
where $C$ is the mean count rate. Then we can deduce the polarization angle $PA = \phi_\mathrm{0} - \pi/2$ and the polarization fraction $\Pi$:
\begin{equation}
\Pi = \frac{a_\mathrm{0}}{a_\mathrm{100}}
\end{equation}
where $a_\mathrm{100}$ represents the amplitude of a 100\,\% polarized source \citep{Suffert1959}. The value of $a_\mathrm{100}$ depends on several factors as the detector dimension, the detection threshold, the level of noise etc. In the case of IBIS, we simulate the emission from a monochromatic source for which we apply the same treatment as for a real source. The resulting modulation is used to deduce the value of $a_\mathrm{100}$ for this energy. We then weight this value by the source spectrum and obtain the value of $a_\mathrm{100}$ for the desired energy band. It is usually around 0.2--0.3 depending on the considered energy band \citep{Laurent2011, Rodriguez2015}. Figure \ref{fig:modulation} shows the evolution of $a_\mathrm{100}$ as a function of the energy.

In order to measure $N(\phi)$, we need to select the simple Compton events, i.e, those for which photons interact only once in ISGRI and once in PICsIT. \textquotedblleft\,Spurious events\,\textquotedblright\ are removed according to the method described in \cite{Forot2007}.
Photons are accumulated in six different angle ranges of 30° each. In order to improve the signal to noise ratio in each channel, we take advantage of the $\pi$ symmetry of the differential cross section described by equation \eqref{eq:cross_section} since e.g., the first channel contains photons with an azimuth 0° $< \phi < $ 30° and photons with an azimuth 180° $< \phi < $ 210°. Shadowgrams are formed for each channel angle chosen by the user, then deconvolved and count rates are extracted. 

The uncertainty on $N(\phi)$ is dominated by statistical fluctuations, since our observations are background dominated. Therefore, confidence intervals for $\phi_0$ and $a_0$ are not derived by a $N(\phi)$ fit to the data but obtained with a Bayesian approach following the work of \cite{Forot2008} and described in \citet{Vaillancourt2006} and \citet{Weisskopf2006}. In this computation, we suppose that all real polarization angles and fractions have a uniform probability distribution \citep[non-informative
prior densities,][]{Quinn2012, Maier2014} and that the
real polarization angle and fraction are $\phi_0$ and $a_0$. We then need
the probability density distribution of measuring $a$ and $\phi$ from $N_\mathrm{pt}$ independent data points in $N(\phi)$ during a period $\pi$, which  is given by \citep{Vaillancourt2006, Forot2008, Maier2014} : 
\begin{multline}
\label{eq:proba_pola}
\mathrm{d}P(a, \phi) = \frac{N_\mathrm{pt} C^2}{\pi \sigma_\mathrm{C}^2} \exp \Bigg[-\frac{N_\mathrm{pt}C^2}{2\sigma_\mathrm{C}^2}  \Bigg. [a^2+a_\mathrm{0}^2  - \\ 2aa_\mathrm{0} \cos(2\phi - 2\phi_\mathrm{0})]  \Bigg.  \Bigg]\, a \, \mathrm{d}a \, \mathrm{d}\phi
\end{multline}
where $\sigma_\mathrm{C}$ is the uncertainty of $C$. Uncertainties of $a$ and $\phi$ can then be deduced by integrating $\mathrm{d}P(a, \phi)$ by respect to the other dimension. We emphasize that this probability is a conditional probability and is calculated by supposing that the emission is indeed polarized.


There are also several systematics uncertainties that arise from measurements of polarization with a Compton telescope using a coded mask. The non-axisymmetric geometry of the detectors and the systematics due to the analysis process have been studied in details in \cite{Forot2008}. We also study the modulation from the background by selecting events from detector pixels hidden from the source by opaque mask elements, for scws of \dixhuit. We find a modulation of 5\,\% for the different energy bands used in the analysis. All these systematics uncertainties are taken into account in the derivation of the polarization constrains measured in this paper.




\begin{table*}[t]
	\center
	\srcsize{
    \caption{Parameters obtained for our polarization analysis using the Compton mode. C and S is the value of the \chired obtained using a constant (C) or a sinusoïdal (S) function for our polarigram fit.}
    \label{tab:polar_analysis}
    \centering
    \begin{tabular}{l c c c c c c c c c c}
    \hline \hline
		 Source & State period & Exposure time & Energy band & Signal & $a_\mathrm{100}$ & $\chi^2$/dof & $p_\mathrm{unpola}$ & Polarization & Polarization & Polarization \\
		& & [Ms] & & to noise ratio & & & [\%] & Angle [°] & fraction [\%] & detected\\
		\hline
		 \quinze & HIMS & 0.22 & 300--400\, keV & 4.9 & 0.278 & C = 2.35/5 & -- & -- & -- & $\times$ \\
		 & & & & & & S = 1.52/4 & & & \\

		 & & & 400--1000\,keV & 7.2 & 0.194 & C = 1.30/5& -- &-- & -- & $\times$ \\
		 & & & & & & S = 0.64/4 & & & \\
		 
		 & & & 300--1000\,keV & 8.7 & 0.224 & C = 1.35/5 & -- & -- & -- & $\times$ \\
		 & & & & & & S = 0.92/4 & & & \\
		 
		  & SIMS & 0.16 & 300--400\, keV & 3.3 & 0.278 & C = 8.00/5 & -- & -- & -- & $\times$ \\
		 & & & & & & S = 5.60/4 & & & \\

		 & & & 400--1000\,keV & 3.7 & 0.194 & C = 2.00/5 & -- & -- & -- & $\times$ \\
		 & & & & & & S = 2.72/4 & & & \\
		 
		 & & & 300--1000\,keV & 4.9 & 0.224 & C = 6.35/5 & -- & -- & -- & $\times$ \\
		 & & & & & & S = 6.40/4 & & & \\
		 \hline
		 \dixhuit & LHS & 1.3 & 300--400\, keV & 71 & 0.278 & C = 12.05/5 & 0.57 & $120 \pm 14$ & $17 \pm 8$ & \checkmark \\
		 & & & & & & S = 6.44/4 & & & \\

		 & & & 400--1000\,keV & 67 & 0.194 & C = 13.1/5 & 0.11 & $105 \pm 11$ & $35 \pm 12$ & \checkmark \\
		 & & & & & & S = 4.48/4 & & & \\
		 
		 & & & 300--1000\,keV & 94 & 0.224 & C = 20.65/5 & 0.003 & $110 \pm 11$ & $26 \pm 9$ & \checkmark \\
		 & & & & & & S = 7.44/4 & & & \\
		 \hline
		 
		 \treize & LHS & 0.32 & 300--400\, keV & 22.9 & 0.278 & C = 12.30/5 & 0.02 & $160 \pm 9$ & $75 \pm 26$ & \checkmark \\
		 & & & & & & S = 0.84/4 & & & \\

		 & & & 400--1000\,keV & 22 & 0.194 & C = 9.40/5 & 0.06 & $180 \pm 10$ & $> 70$ & \checkmark \\
		 & & & & & & S = 5.28/4 & & & \\
		 
		 & & & 300--1000\,keV & 30.7 & 0.224 & C = 15.45/5 & 0.008 & $180 \pm 10$ & $79 \pm 23$ & \checkmark \\
		 & & & & & & S = 0.68/4 & & & \\
		 
		 & IMS & 0.17 & 300--400\, keV & 8.2 & 0.278 & C = 13.15/5 & -- & -- & -- & $\times$ \\
		 & & & & & & S = 12.84/4 & & & \\

		 & & & 400--1000\,keV & 7.2 & 0.194 & C = 4.95/5 & -- & -- & -- & $\times$ \\
		 & & & & & & S = 3.88/4 & & & \\
		 
		 & & & 300--1000\,keV & 10.42 & 0.224 & C = 7.45/5 & -- & -- & -- & $\times$ \\
		 & & & & & & S = 6.32/4 & & & \\
		 
		 & HSS & 0.42 & 300--400\, keV & 2.79 & 0.278 & C = 3.15/5 & -- & -- & -- & $\times$ \\
		 & & & & & & S = 0.48/4 & & & \\

		 & & & 400--1000\,keV & 3.01 & 0.194 & C = 11.01/5 & -- & -- & -- & $\times$ \\
		 & & & & & & S = 4.08/4 & & & \\
		 
		 & & & 300--1000\,keV & 4.12 & 0.224 & C = 5.55/5 & -- & -- & -- & $\times$ \\
		 & & & & & & S = 1.32/4 & & & \\

		 \hline
		 
        \hline
    \end{tabular}
    }
\end{table*}

\begin{figure*}[h]
    \centering
    \includegraphics[width=\textwidth]{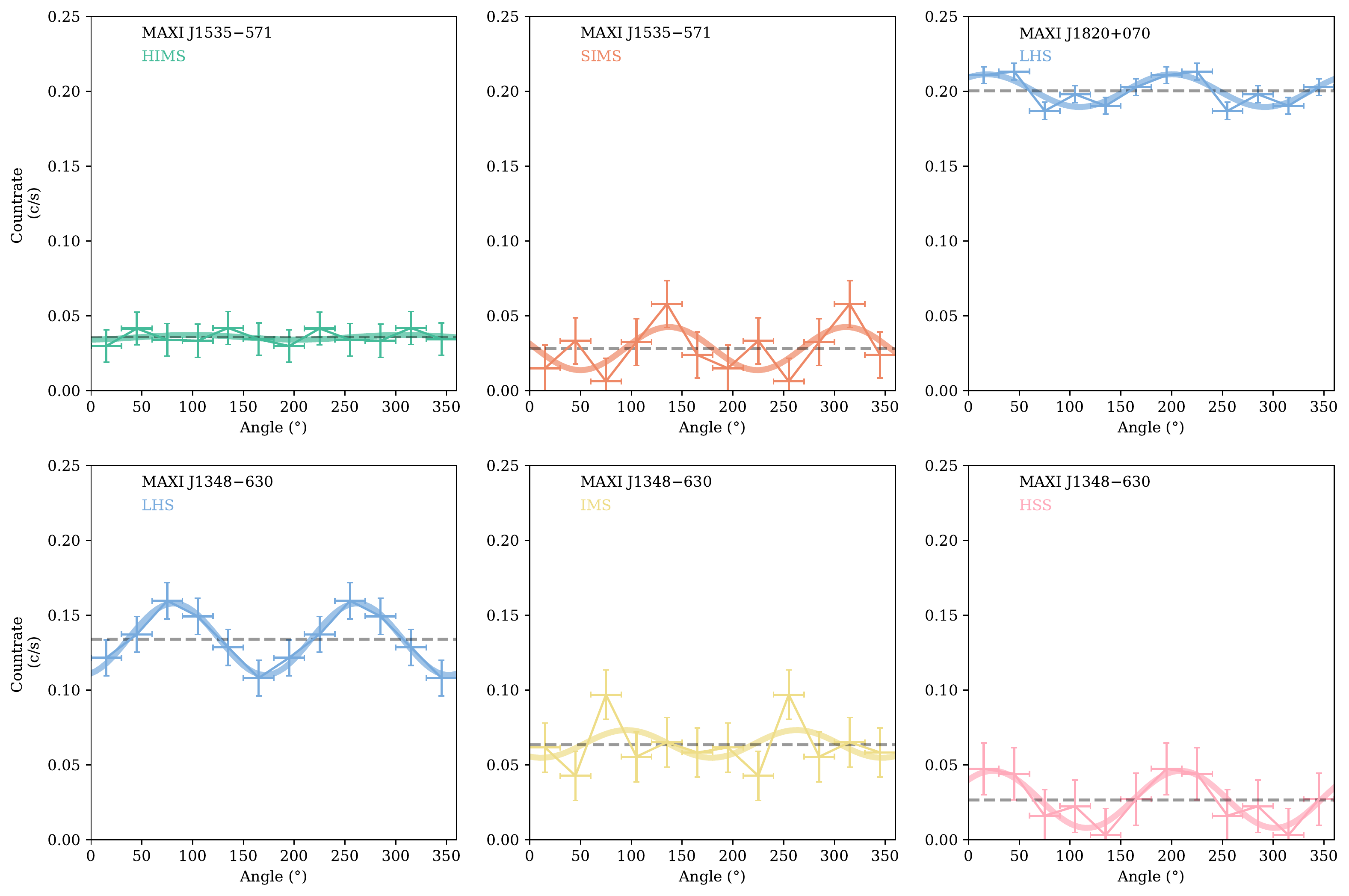}
    \caption{Polarigrams of the three sources obtained in the 300--1000\,keV energy band range.}
    \label{fig:pola}
\end{figure*}

\subsection{Results}

\begin{figure*}[h]
    \centering
    \includegraphics[width=\textwidth]{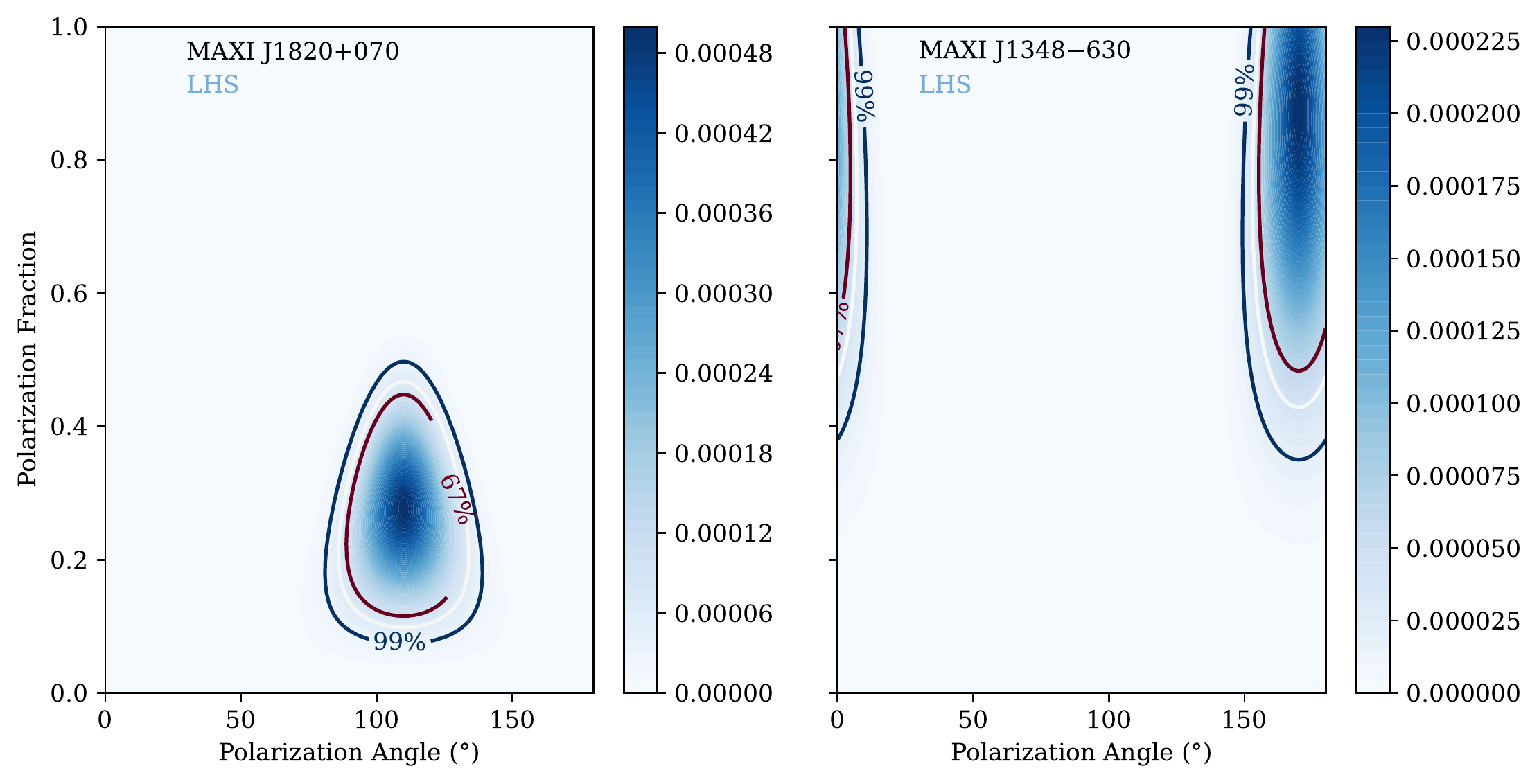}
    \caption{Probability density described by equation \eqref{eq:proba_pola} in function of the polarization angle and the polarization fraction calculated for \dixhuit (left) and \treize in the LHS (right).}
    \label{fig:proba_pola}
\end{figure*}

Figure \ref{fig:pola} shows polarigrams in the 300--1000\,keV band for the different periods of the three sources, following the same color code as in Fig. \ref{fig:lightcurves}. Figures \ref{fig:pola_300-400} and \ref{fig:pola_400-1000} show polarigrams in the 300--1000\,keV and 400--1000\,keV bands respectively. Polarigrams are fitted with a constant (dashed grey line) and the sinusoid function described in equation \eqref{eq:polarigram} (colored line). During the LHS periods of \dixhuit and \treize, we also show in Fig. \ref{fig:proba_pola} the integrated probability function described by equation \eqref{eq:proba_pola} in function of the polarization angle $\phi$ and the polarization fraction $\Pi$. Other probability density functions are shown on Fig. \ref{fig:all_contours}. Table \ref{tab:polar_analysis} summarizes the different parameters we measure for the three sources and for three different energy bands: 300--400\,keV, 400--1000\,keV and 300--1000\,keV. We indicate the total effective exposure time, the signal to noise ratio, the $a_\mathrm{100}$ value, the \chired obtained by fitting $N(\phi)$ by a constant (C) or by the sinusoid function (S) described by equation \eqref{eq:polarigram}, the polarization angle $PA$ and the polarization fraction $\Pi$. Note that we use a range of 0--180° for our fitting. Uncertainties represent an interval confidence of 67\,\%. The last column indicates whether polarization is detected (\checkmark) or not ($\times$). We consider that polarization is detected by validating two conditions: 
(1) we need a signal to noise ratio higher than 12 to obtain reliable results; this value is based on empirical results on the Crab \citep{Laurent2016}, and (2) the probability $p_\mathrm{unpola}$ of measuring modulation knowing that the source is unpolarized is $< 1$\,\%. All $p_\mathrm{unpola}$ are shown in Table \ref{tab:polar_analysis} (only for polarigrams where the signal to noise ration is higher than 12).



We do not detect polarization for \quinze. Indeed, for both states and for the three energy bands, the signal to noise ratio is too poor and no modulation  is detected.


The diagnosis is different for \dixhuit. Indeed, for the three considered energy bands, polarigrams show clear deviation from a constant which poorly represents the data 
We find a much better description of the data with the sine function (eq. \ref{eq:polarigram}). In the three energy bands, we calculate the probability given by equation (eq. \ref{eq:proba_pola}) using the values of $a$ and $\phi$ we find in our best fit. Figure \ref{fig:proba_pola} (left) shows the contour plot we obtain in the 300--1000\,keV bands; polarization is detected with a interval confidence higher than 99\,\%, and we find a polarization angle and a polarization fraction consistent with $PA \sim 110$° and $\Pi \sim 25$\,\% in the three energy bands.

Concerning \treize, polarigrams extracted from the IMS and HSS periods have poor signal to noise ratio. On the other hand, polarigrams from the LHS period have sufficient signal to noise to probe the presence of a modulation. We find that polarigrams are poorly described by a constant and $p_\mathrm{unpol} < 1$\,\% in the different energy bands. We then calculate the probability (eq. \ref{eq:proba_pola}) to measure the polarization angle $PA$ and a polarization fraction $\Pi$. The contour plot is shown on Fig. \ref{fig:proba_pola} (right). We find lower limits for the polarization fraction of 49\,\%, 70\,\%and 56\,\% with a polarization angle of $\sim 160$\,\textdegree, $\sim 180$\,\textdegree\  and $\sim 180$\,\textdegree\ for the 300--400\,keV, 300--1000\,keV and 300--1000\,keV bands respectively.

\section{Discussion and interpretation}
\label{sec:interpretation}

\subsection{Summary of the results}
\subsubsection{\quinze}

We separate the data in two intervals corresponding to a HIMS and SIMS, which respectively match epochs defined in \citet{Russell2020}. In both periods, the source spectra are rather soft and characterized by a photon index $\Gammath^\mathrm{HIMS} = 2.38^{+0.05}_{-0.09}$ and $\Gammath^\mathrm{SIMS} = 2.52^{+0.08}_{-0.09}$. The low energy emission starts to be dominated by the accretion disk during the SIMS period and its energy temperature peaks at $\kTz = 1.24^{+0.04}_{-0.03}$\,keV which is a typical value usually observed in soft states of BHBs \citep[e.g.,][]{Remillard2006}. The energy from the corona electrons is not well defined in both states with a value greater than 50\,keV. 

We detect an additional component above the Comptonization bump during both states. This high-energy component is described by a powerlaw with a photon index of $\Gammapo^\mathrm{HIMS} = 2.1^{+0.4}_{-0.5}$ and $\Gammapo^\mathrm{SIMS} = 2.0^{+0.4}_{-0.4}$.

Using the Compton mode, we do not detect polarization in either both states. We remark, however, that the source flux above 300\,keV is quite low (see Table \ref{tab:spectral_analysis}) resulting in a poor signal to noise ratio for our polarization measurements. The non-detection is compatible with this empirical minimum value to obtain a trustworthy detection of signal. Therefore, we cannot conclude for \quinze if the lack of polarization detection at high energy is an intrinsically characterictic of the source or an observationnal issue.





\subsubsection{\dixhuit}

\dixhuit is the brightest of the three sources observed in the ISGRI 30--50\,keV range. The source is observed during its LHS \citep[e.g.,][]{Buisson2019}. The spectrum is characterized by a hard photon index \Gammath $=1.61^{+0.01}_{-0.01}$ and corona electron energy of $57^{+4}_{-4}$\,keV. An additional powerlaw component clearly improves the fit and we find a photon index of \Gammapo $=2.09^{+0.03}_{-0.03}$. This component strongly dominates the spectrum above 300\,keV with a flux of $4.6 \times 10^{-9}$\,erg\,s$^{-1}$\,cm$^{-2}$. 

Polarigrams extracted in the 300--400\,keV, 400--1000\,keV and 300--1000\,keV show a strong modulation of the signal and we measure a 300--1000\,keV polarization fraction of $\Pi = 26 \pm 9$ and a polarization angle $PA = 110 \pm 11$° (see Tab. \ref{tab:polar_analysis} for the energy dependent results), errors are given in a 67\,\% confidence range. 


\subsubsection{\treize}

We follow the evolution of the source during its outburst and identify three different periods corresponding to three different states of the source, LHS, IMS and HSS. During its LHS, we find a photon index of \Gammath = $1.69^{+0.04}_{-0.03}$ and an electron energy $\kT = 44^{+5}_{-4}$\,keV. We detect an additional powerlaw component with a photon index of $\Gammapo = 2.0^{+0.3}_{+0.3}$. 


While a high-energy tail is clearly present in the LHS, no high-energy component is detected in either of the softer states. Consistently polarization can be probed only in the LHS, and we indeed detect the same polarization angle $PA = 180 \pm 10$° for the 400--1000\,keV and 300--1000\,keV bands. This value is not strictly consistent with the value of $PA = 160 \pm 9$° we measure in the 300--400\,keV band, but it is consistent at 90\,\% confidence ($PA = 160 \pm 30$°). This slight discrepancy could arise from some contribution of the Comptonized continuum, as the 300--400 keV range is not purely described by the additional powerlaw. We measure a high polarization fraction for the three energy bands with a lower limit $> 70$\,\% for the 400–1000 keV. The other energy bands allows lower polarization fractions; $\Pi^{300-400 keV} > 49$\,\% and $\Pi^{300-1000 keV} > 56$\,\%.

\subsection{Origin of the high-energy emission}

We detect a powerlaw tail in addition to the standard Comptonisation component during the HIMS and SIMS of \quinze, the LHS of \dixhuit and during the LHS of \treize. This component is strongly present at the beginning of the outburst when the global spectral shape is hard and its strenght decays as the outburst evolves to softer states. There are two obvious possibilities for the \textquotedblleft apparent\textquotedblright\ absence of high-energy tails in softer states. Either 1) the emission genuinely vanishes; 2) the high-energy emission falls below detection threshold. 

Case 1) implies that the medium responsible for the high-energy component disappears in the softer states. One obvious candidate is the compact jet, which is known to be quenched in the HSS \citep[e.g.,][]{Fender1999}, and was claimed to emit in the 400--2000 keV range in Cygnus X-1 \citep{Laurent2011, Jourdain2012, Rodriguez2015}. Since ejections of coronae state transitions have been proposed in other sources \citep[e.g.,][]{Rodriguez2008a, Rodriguez2008b}, the corona would also be a good candidate.
Case 2) implies an evolution of the high-energy component parameters (at least its flux) which also leads to at least two interpretations. 2a) Same medium with evolving properties, e.g., in Cygnus X-3, the tails seems un-related to the jets behavior, and well explained by an hybrid corona in all states \citep{Cangemi2021a}. 2b) Different emitting media, as observed in Cygnus X-1 \citep{Cangemi2021b}. In the case of Cygnus X-1, we note that the jets origin in both states could also be a possibility \citep{Zdziarski2020}.

Polarization is an additional diagnostic to constrain the origin of the tail. The synchrotron spectrum of a population of $N$ electrons with an energy between $E$ and $E+dE$ and with an electron index $p$, $dN(E) \propto E^{-p}dE$ can be approximated by a powerlaw $P(E) \propto E^{-\alpha}$. The spectral index $\alpha$ and the electron index $p$ are tied by the equation $\alpha = (p-1)/2$ \citep{Rybicki&Lightman1986} and $\alpha$ can also be related to the photon index $\Gamma$ : $\alpha = \Gamma - 1$. In a very ordered magnetic field, the polarization fraction of a polarized emission expected in the optically thin regime of the synchrotron spectrum, is \citep{Rybicki&Lightman1986}:
\begin{equation}
\Pi = \frac{p + 1}{p + \frac{7}{3}}
\label{eq:pola}
\end{equation}

In the case of \treize, the value of the photon index during the LHS $\Gammapo = 2.0^{+0.3}_{-0.3}$ leads to a polarization fraction of $\Pi = 75^{+11}_{-11}$\,\% which is consistent with the polarization fraction observed in the three energy bands considered. Therefore, the observed emission could arise from the synchrotron emission from the jets base in a very ordered magnetic field. Besides, synchrotron emission from compact radio jets has been observed during the LHS of \treize \citep[e.g.,][]{Russell2020, Carotenuto2021}. 

While the upper limit (or non-detection) in the case of \quinze does not allow to conclude, the case of \dixhuit is quite interesting. The rather low level of polarization (compared to \treize and Cygnus X-1) may indicate various possibilities. Here the measured value of the photon index $\Gammapo = 2.09^{+0.03}_{-0.03}$ leads to a polarization fraction of $75 \pm 2$\,\% not consistent with the polarization fraction we measure. This could indicates (1) the magnetic field is disrupted by some mechanism and/or differents jet zones emits polarized radiation with different angles; or (2) another origin than the jets for the polarized emission.



\subsection{SEDs basic analysis} 

While a precise spectral modelisation with physical models is beyond the scope of this paper, we can apply a simple approach using spectral energy distribution to verify the jet hypothesis. Here, the goal is to investigate whether the optically thin part of the synchrotron spectrum from the jets is consistent with our measured high-energy component. Therefore, we need the information on the synchrotron break frequency of the jets spectrum as well as the spectral index of the optically thin part. Then, we can extrapolate the synchrotron spectrum up to 1\,MeV and check the consistency with the hard X-rays. 

\subsubsection{\quinze}
\begin{figure}[h]
\includegraphics[width=\columnwidth]{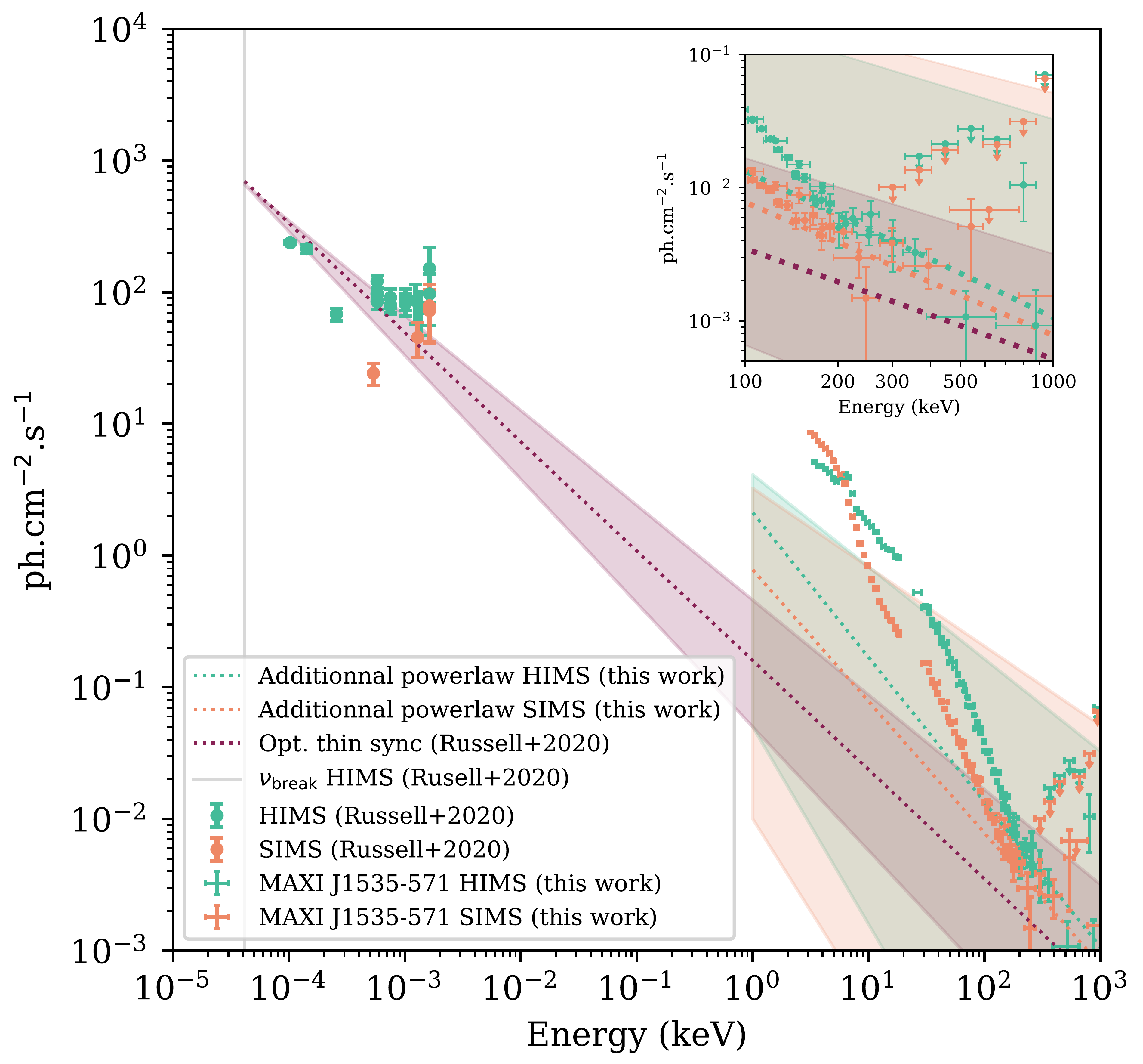}
\caption{\quinze spectra using observational data from \cite{Russell2020} and this work. HIMS and SIMS data are represented in green and orange, respectively. The measured additional high-energy components are indicated with dotted line. Their  90\,\% uncertainties interval range are represented by the colored area. The purple dotted line is the extrapolation of the optically thin synchrotron spectrum from \cite{Russell2020} with $\alpha = 0.83 \pm 0.09$.}
\label{fig:1535_broad}
\end{figure}

Figure \ref{fig:1535_broad} shows different datasets from \cite{Russell2020} in the HIMS (green) and in the SIMS (orange). The grey line indicates the energy of the synchrotron break, whereas the green line is the optically thin part of the synchrotron spectrum obtained by \cite{Russell2020} ($\alpha = 0.83 \pm 0.09$, where $\alpha$ is the spectral index of the optically thin part of the synchrotron spectrum), 
extrapolated up to 1000\,keV. We also show the spectra extracted in this work and their associated measured powerlaws. The box on the top right corner of the figure is an enlarged view from 100\,keV to 1000\,keV. We observe that the extrapolation of the optically thin spectrum is consistent, within the errorbars, with the powerlaw in the HIMS. This could point towards a jets origin of the high-energy component in this state. However, the uncertainties are large at high energy, and without polarization measurements, we cannot exclude a hybrid corona origin. Concerning the SIMS, \cite{Russell2020} shows that the radio emission is quenched in this state. Therefore, the non-thermal component $>200$\,keV in this state could come from another region from the jets that still radiates in the X-ray or has another origin. 

\subsubsection{\dixhuit}

\begin{figure}[h]
\includegraphics[width=\columnwidth]{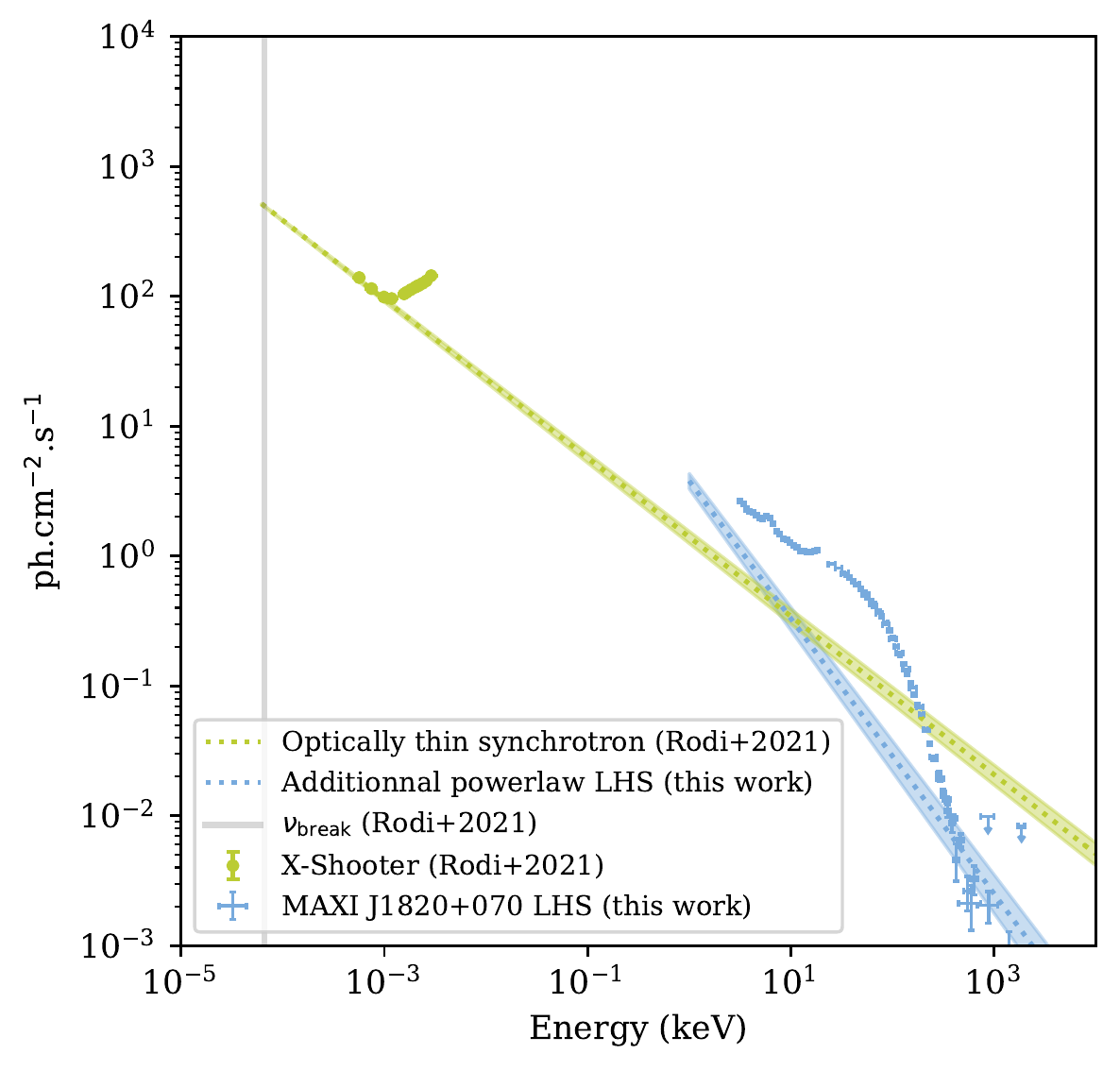}
\caption{\dixhuit spectrum using observational data from \cite{Rodi2021} and this work. The measured additional high-energy component is indicated with the blue dotted line. The green dotted line is the extrapolation of the optically thin synchrotron spectrum from \cite{Rodi2021}.}
\label{fig:1820_broad}
\end{figure}

The dotted green line on Fig. \ref{fig:1820_broad} shows the extrapolation of the optically thin synchrotron spectrum measured with simulatenous (12th of april 2018) \textit{X-Shooter} data \citep[green dots][]{Rodi2021}. The X-ray spectrum from this work is shown with blue dots whereas the dotted blue line shows our measured powerlaw. Here, the IR optically thin spectrum is not consistent with the high-energy component that we measure. This result clearly excludes a pure synchrotron jets origin for the additional component and favours a hybrid corona origin as proposed by \cite{Zdziarski2021}. Regarding our polarization measurements, a polarization fraction of $26 \pm 9$\,\% is also consistent with a hybrid corona origin \citep{Beheshtipour2017}.

\subsubsection{\treize}

\begin{figure}[h]
\includegraphics[width=\columnwidth]{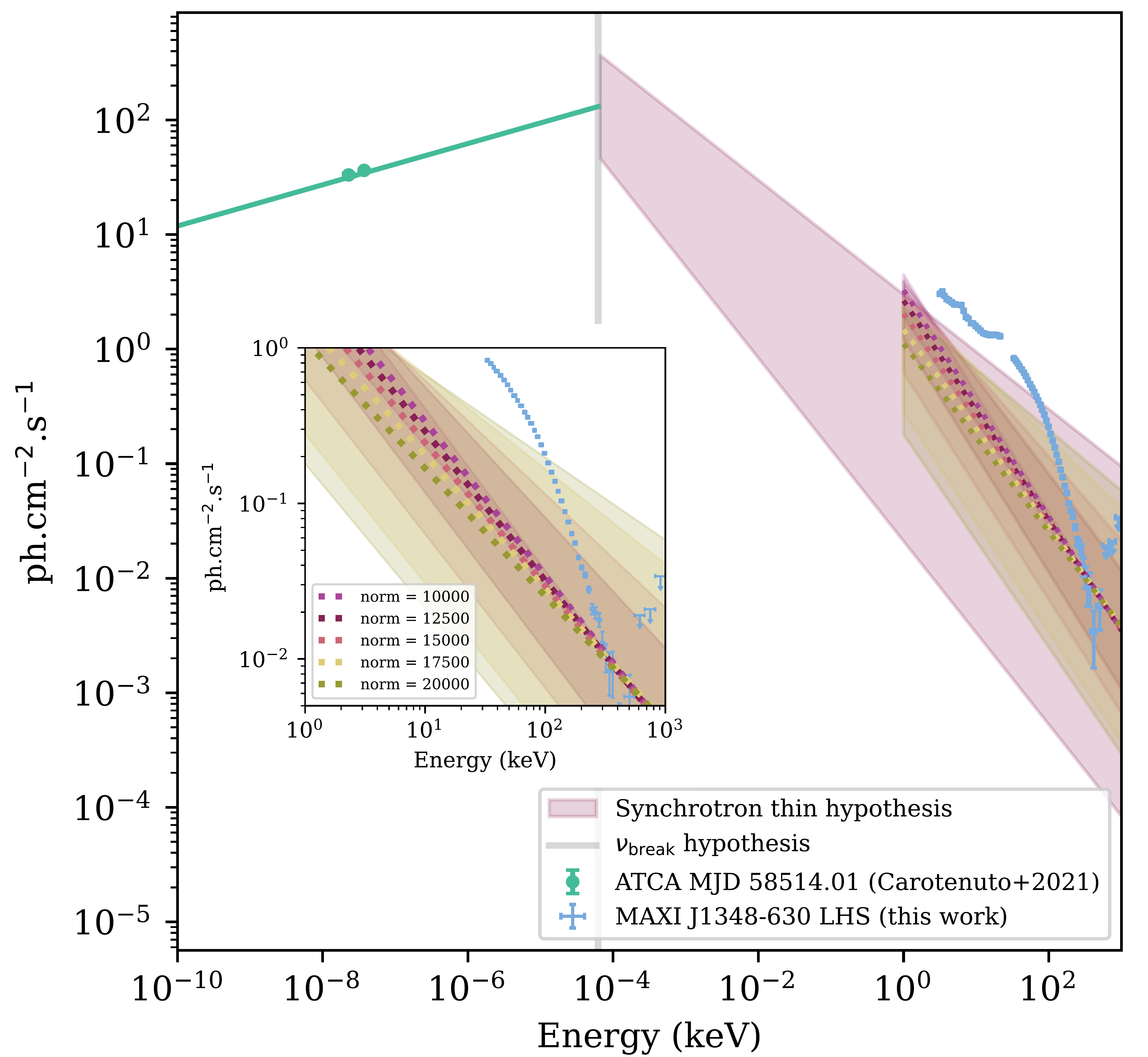}
\caption{Broad band spectrum of \treize. The light blue line is the spectrum as measured by \cite{Carotenuto2021}. The green zone corresponds to our synchrotron thin spectrum hypothesis (see the text). The insert plot focus on the 1--1000\,keV range. The different colors indicates the resulting powerlaw for different normalization value of the \textsc{diskbb}.}
\label{fig:1348_broad}
\end{figure}

No infrared data has been published on \treize so far. Therefore, we do not have informations on the synchrotron cutoff. However, we try to investigate the consistency of the high-energy emission detected with synchrotron emission from the jets assuming a spectral index $-0.5 < \alpha < 0.8$, a flux $40 < \mathrm{F} < 300$\,mJy at the synchrotron cutoff energy $1.5 \times 10^{15}$\,Hz. Figure \ref{fig:1348_broad} shows the resulting broad band spectrum. We also plot the optically thick part of the synchrotron spectrum as measured by \cite{Carotenuto2021} with \textit{ATCA} at MJD 58514.01.

As we add a disk to model the spectrum (see Sect. \ref{sec:spectral}), we investigate the impact of the black body disk normalization value on the powerlaw normalization and photon index values. We let the normalization parameter of \textsc{diskbb} vary from $1 \times 10^{4}$ to $2 \times 10^{4}$ which is the upper limit obtained by \cite{Chakraborty2020}. The resulting powerlaws from the fit are shown in different colors on Fig. \ref{fig:1348_broad} and the insert is an enlarged view from 1\,keV to 1000\,keV. The normalization of the black body disk has a strong impact on the powerlaw properties and we note that for normalization higher than $\sim 15000$, the powerlaw component is consistent with the purple zone and therefore with the synchrotron scenario. However, although the high polarization fraction is also consistent with synchrotron emission, our synchrotron hypothesis also strongly depends on the energy of the synchrotron break, and we cannot formally conclude for a synchrotron origin without any precise measurement of the optically thin part of the synchrotron spectrum.

\section{Summary and conclusion}

In this work, we use the \integral unique capabilities to investigate the high-energy properties of three sources, \quinze, \dixhuit, \treize during their outburst. For each outburst, we divide the data into different state periods based on their spectral characteristics. Thanks to the combination of JEM-X, ISGRI and SPI data, we create stacked spectra in the 3--2000\,keV band range for each of these states. We then use a simple phenomenological spectral approach in order to investigate the behavior of the sources at high-energy and search for an additional high-energy component. 

We use the Compton mode of \integral/IBIS to study polarization properties and find that the 300--1000\,keV emission from the LHS of \dixhuit and \treize are polarized.


We extrapolate the optically thin part observed in the IR of the synchrotron spectrum to investigate the potential origin of the high-energy component detected in these sources. In the HIMS of \quinze, the synchrotron spectrum is consistent with the detected high-energy component. However, we could not measure polarization in this source and therefore cannot exclude a hybrid corona origin. For \dixhuit, the extrapolation of the synchrotron spectrum is not consistent with the extension measured in the X-rays. Therefore the non-thermal component could arise from a non-thermal distribution of electron from the corona. The polarization fraction that we measure is also consistent with this scenario. In the case of the high-energy component detected in the LHS of \treize, the high polarization fraction we measure is consistent with synchrotron emission in a very ordered magnetic field. However, we prefer not to conclude on the origin of the high-energy component without any clear information on the optically thin part of the synchrotron spectrum.




\begin{acknowledgements}
J.R., P.L. P.-O.P \& C.G. acknowledge partial funding from the French Space Agency (CNES). J.R. \& P.-O.P acknowledge partial fundings from the French Programme National des Hautes Energies (PNHE). T.M.B. acknowledges financial contribution from the agreement ASI-INAF n. 2017-14-H.0 and from PRIN INAF 2019 n.15. Based on observations with \integral, an ESA project with instruments and science data centre funded by ESA member states (especially the PI countries: Denmark, France, Germany, Italy, Switzerland, Spain) and with the participation of Russia and the USA.

\end{acknowledgements}

\bibliographystyle{aa}
\bibliography{pola}

\begin{thebibliography}{91}
\expandafter\ifx\csname natexlab\endcsname\relax\def\natexlab#1{#1}\fi

\bibitem[{{Anczarski} {et~al.}(2020){Anczarski}, {Neilsen}, {Remillard},
  {Uttley}, {Arzoumanian}, {Gendreau}, \& {Steiner}}]{Anczarski2020}
{Anczarski}, J., {Neilsen}, J., {Remillard}, R., {et~al.} 2020, in American
  Astronomical Society Meeting Abstracts, Vol. 235, American Astronomical
  Society Meeting Abstracts \#235, 369.02

\bibitem[{{Anders} \& {Ebihara}(1982)}]{AndersEbihara1982}
{Anders}, E. \& {Ebihara}, M. 1982, Meteoritics, 17, 180

\bibitem[{{Arnaud}(1996)}]{Arnaud1996}
{Arnaud}, K.~A. 1996, in Astronomical Society of the Pacific Conference Series,
  Vol. 101, Astronomical Data Analysis Software and Systems V, ed. G.~H.
  {Jacoby} \& J.~{Barnes}, 17

\bibitem[{{Atri} {et~al.}(2020){Atri}, {Miller-Jones}, {Bahramian}, {Plotkin},
  {Deller}, {Jonker}, {Maccarone}, {Sivakoff}, {Soria}, {Altamirano},
  {Belloni}, {Fender}, {Koerding}, {Maitra}, {Markoff}, {Migliari}, {Russell},
  {Russell}, {Sarazin}, {Tetarenko}, \& {Tudose}}]{Atri2020}
{Atri}, P., {Miller-Jones}, J.~C.~A., {Bahramian}, A., {et~al.} 2020, \mnras,
  493, L81

\bibitem[{{Beheshtipour} {et~al.}(2017){Beheshtipour}, {Krawczynski}, \&
  {Malzac}}]{Beheshtipour2017}
{Beheshtipour}, B., {Krawczynski}, H., \& {Malzac}, J. 2017, \apj, 850, 14

\bibitem[{{Belloni}(2010)}]{Belloni2010}
{Belloni}, T.~M. 2010, {States and Transitions in Black Hole Binaries}, ed.
  T.~{Belloni}, Vol. 794, 53

\bibitem[{{Belloni} {et~al.}(2020){Belloni}, {Zhang}, {Kylafis}, {Reig}, \&
  {Altamirano}}]{Belloni2020}
{Belloni}, T.~M., {Zhang}, L., {Kylafis}, N.~D., {Reig}, P., \& {Altamirano},
  D. 2020, \mnras, 496, 4366

\bibitem[{{Bhargava} {et~al.}(2019){Bhargava}, {Belloni}, {Bhattacharya}, \&
  {Misra}}]{Bhargava2019}
{Bhargava}, Y., {Belloni}, T., {Bhattacharya}, D., \& {Misra}, R. 2019, \mnras,
  488, 720

\bibitem[{{Bozzo} {et~al.}(2018){Bozzo}, {Savchenko}, {Ferrigno}, {Ducci},
  {Kuulkers}, {Ubertini}, \& {Laurent}}]{Bozzo2018}
{Bozzo}, E., {Savchenko}, V., {Ferrigno}, C., {et~al.} 2018, The Astronomer's
  Telegram, 11478, 1

\bibitem[{{Bright} {et~al.}(2020){Bright}, {Fender}, {Motta}, {Williams},
  {Moldon}, {Plotkin}, {Miller-Jones}, {Heywood}, {Tremou}, {Beswick},
  {Sivakoff}, {Corbel}, {Buckley}, {Homan}, {Gallo}, {Tetarenko}, {Russell},
  {Green}, {Titterington}, {Woudt}, {Armstrong}, {Groot}, {Horesh}, {van der
  Horst}, {K{\"o}rding}, {McBride}, {Rowlinson}, \& {Wijers}}]{Bright2020}
{Bright}, J.~S., {Fender}, R.~P., {Motta}, S.~E., {et~al.} 2020, Nature
  Astronomy, 4, 697

\bibitem[{{Buisson} {et~al.}(2019){Buisson}, {Fabian}, {Barret}, {F{\"u}rst},
  {Gandhi}, {Garc{\'\i}a}, {Kara}, {Madsen}, {Miller}, {Parker}, {Shaw},
  {Tomsick}, \& {Walton}}]{Buisson2019}
{Buisson}, D.~J.~K., {Fabian}, A.~C., {Barret}, D., {et~al.} 2019, \mnras, 490,
  1350

\bibitem[{{Cadolle Bel} {et~al.}(2006){Cadolle Bel}, {Sizun}, {Goldwurm},
  {Rodriguez}, {Laurent}, {Zdziarski}, {Foschini}, {Goldoni}, {Gouiff{\`e}s},
  {Malzac}, {Jourdain}, \& {Roques}}]{CadolleBel2006}
{Cadolle Bel}, M., {Sizun}, P., {Goldwurm}, A., {et~al.} 2006, \aap, 446, 591

\bibitem[{{Cangemi} {et~al.}(2019{\natexlab{a}}){Cangemi}, {Belloni}, \&
  {Rodriguez}}]{Cangemi2019b}
{Cangemi}, F., {Belloni}, T., \& {Rodriguez}, J. 2019{\natexlab{a}}, The
  Astronomer's Telegram, 12471, 1

\bibitem[{{Cangemi} {et~al.}(2021{\natexlab{a}}){Cangemi}, {Beuchert},
  {Siegert}, {Rodriguez}, {Grinberg}, {Belmont}, {Gouiff{\`e}s}, {Kreykenbohm},
  {Laurent}, {Pottschmidt}, \& {Wilms}}]{Cangemi2021b}
{Cangemi}, F., {Beuchert}, T., {Siegert}, T., {et~al.} 2021{\natexlab{a}},
  \aap, 650, A93

\bibitem[{{Cangemi} {et~al.}(2019{\natexlab{b}}){Cangemi}, {Rodriguez},
  {Belloni}, {Clavel}, \& {Grinberg}}]{Cangemi2019a}
{Cangemi}, F., {Rodriguez}, J., {Belloni}, T., {Clavel}, M., \& {Grinberg}, V.
  2019{\natexlab{b}}, The Astronomer's Telegram, 12457, 1

\bibitem[{{Cangemi} {et~al.}(2021{\natexlab{b}}){Cangemi}, {Rodriguez},
  {Grinberg}, {Belmont}, {Laurent}, \& {Wilms}}]{Cangemi2021a}
{Cangemi}, F., {Rodriguez}, J., {Grinberg}, V., {et~al.} 2021{\natexlab{b}},
  \aap, 645, A60

\bibitem[{{Carotenuto} {et~al.}(2021){Carotenuto}, {Corbel}, {Tremou},
  {Russell}, {Tzioumis}, {Fender}, {Woudt}, {Motta}, {Miller-Jones}, {Chauhan},
  {Tetarenko}, {Sivakoff}, {Heywood}, {Horesh}, {van der Horst}, {Koerding}, \&
  {Mooley}}]{Carotenuto2021}
{Carotenuto}, F., {Corbel}, S., {Tremou}, E., {et~al.} 2021, \mnras, 504, 444

\bibitem[{{Chakraborty} {et~al.}(2020){Chakraborty}, {Navale}, {Ratheesh}, \&
  {Bhattacharyya}}]{Chakraborty2020}
{Chakraborty}, S., {Navale}, N., {Ratheesh}, A., \& {Bhattacharyya}, S. 2020,
  \mnras, 498, 5873

\bibitem[{{Chauhan} {et~al.}(2019){Chauhan}, {Miller-Jones}, {Anderson},
  {Raja}, {Bahramian}, {Hotan}, {Indermuehle}, {Whiting}, {Allison},
  {Anderson}, {Bunton}, {Koribalski}, \& {Mahony}}]{Chauhan2019}
{Chauhan}, J., {Miller-Jones}, J.~C.~A., {Anderson}, G.~E., {et~al.} 2019,
  \mnras, 488, L129

\bibitem[{{Corbel} {et~al.}(2013){Corbel}, {Coriat}, {Brocksopp}, {Tzioumis},
  {Fender}, {Tomsick}, {Buxton}, \& {Bailyn}}]{Corbel2013}
{Corbel}, S., {Coriat}, M., {Brocksopp}, C., {et~al.} 2013, \mnras, 428, 2500

\bibitem[{{Corbel} {et~al.}(2001){Corbel}, {Kaaret}, {Jain}, {Bailyn},
  {Fender}, {Tomsick}, {Kalemci}, {McIntyre}, {Campbell-Wilson}, {Miller}, \&
  {McCollough}}]{Corbel2001}
{Corbel}, S., {Kaaret}, P., {Jain}, R.~K., {et~al.} 2001, \apj, 554, 43

\bibitem[{{Del Santo} {et~al.}(2013){Del Santo}, {Malzac}, {Belmont},
  {Bouchet}, \& {De Cesare}}]{DelSanto2013}
{Del Santo}, M., {Malzac}, J., {Belmont}, R., {Bouchet}, L., \& {De Cesare}, G.
  2013, \mnras, 430, 209

\bibitem[{{Din{\c{c}}er}(2017)}]{Dincer2017}
{Din{\c{c}}er}, T. 2017, The Astronomer's Telegram, 10716, 1

\bibitem[{{Dzie{\l}ak} {et~al.}(2021){Dzie{\l}ak}, {De Marco}, \&
  {Zdziarski}}]{Dzielak2021}
{Dzie{\l}ak}, M.~A., {De Marco}, B., \& {Zdziarski}, A.~A. 2021, \mnras, 506,
  2020

\bibitem[{{Evans} \& {Beiser}(1956)}]{Evans&Beiser1956}
{Evans}, R.~D. \& {Beiser}, A. 1956, Physics Today, 9, 33

\bibitem[{{Fender} {et~al.}(1999){Fender}, {Corbel}, {Tzioumis}, {McIntyre},
  {Campbell-Wilson}, {Nowak}, {Sood}, {Hunstead}, {Harmon}, {Durouchoux}, \&
  {Heindl}}]{Fender1999}
{Fender}, R., {Corbel}, S., {Tzioumis}, T., {et~al.} 1999, \apjl, 519, L165

\bibitem[{{Forot} {et~al.}(2008){Forot}, {Laurent}, {Grenier}, {Gouiff{\`e}s},
  \& {Lebrun}}]{Forot2008}
{Forot}, M., {Laurent}, P., {Grenier}, I.~A., {Gouiff{\`e}s}, C., \& {Lebrun},
  F. 2008, \apjl, 688, L29

\bibitem[{{Forot} {et~al.}(2007){Forot}, {Laurent}, {Lebrun}, \&
  {Limousin}}]{Forot2007}
{Forot}, M., {Laurent}, P., {Lebrun}, F., \& {Limousin}, O. 2007, \apj, 668,
  1259

\bibitem[{{Fuchs} {et~al.}(2003){Fuchs}, {Rodriguez}, {Mirabel}, {Chaty},
  {Rib{\'o}}, {Dhawan}, {Goldoni}, {Sizun}, {Pooley}, {Zdziarski},
  {Hannikainen}, {Kretschmar}, {Cordier}, \& {Lund}}]{Fuchs2003}
{Fuchs}, Y., {Rodriguez}, J., {Mirabel}, I.~F., {et~al.} 2003, \aap, 409, L35

\bibitem[{{Grove} {et~al.}(1998){Grove}, {Johnson}, {Kroeger}, {McNaron-Brown},
  {Skibo}, \& {Phlips}}]{Grove1998}
{Grove}, J.~E., {Johnson}, W.~N., {Kroeger}, R.~A., {et~al.} 1998, \apj, 500,
  899

\bibitem[{{Hannikainen} {et~al.}(1999){Hannikainen}, {Hunstead}, {Durouchoux},
  {Vilhu}, {Campbell-Wilson}, \& {Williamson}}]{Hannikainen1999}
{Hannikainen}, C.~D., {Hunstead}, W.~R., {Durouchoux}, P., {et~al.} 1999,
  Astrophysical Letters and Communications, 38, 237

\bibitem[{{Hoang} {et~al.}(2019){Hoang}, {Molina}, {Lopez}, {Rib{\'o}},
  {Blanch}, {Cortina}, {Maier}, {Park}, {De Naurois}, {de O{\~n}a Wilhelmi},
  {Ernenwein}, {Malyshev}, {Mitchell}, {Ohm}, \& {Zanin}}]{Hoang2019}
{Hoang}, J., {Molina}, E., {Lopez}, M., {et~al.} 2019, in International Cosmic
  Ray Conference, Vol.~36, 36th International Cosmic Ray Conference (ICRC2019),
  696

\bibitem[{{Huang} {et~al.}(2018){Huang}, {Qu}, {Zhang}, {Bu}, {Chen}, {Tao},
  {Zhang}, {Lu}, {Li}, {Song}, {Xu}, {Cao}, {Chen}, {Liu}, {Chang}, {Yu},
  {Weng}, {Hou}, {Kong}, {Xie}, {Zhang}, {ZHOU}, {Chang}, {Chen}, {Chen},
  {Chen}, {Chen}, {Cui}, {Cui}, {Deng}, {Dong}, {Du}, {Fu}, {Gao}, {Gao},
  {Gao}, {Ge}, {Gu}, {Guan}, {Gungor}, {Guo}, {Han}, {Hu}, {Huo}, {Ji}, {Jia},
  {Jiang}, {Jiang}, {Jin}, {Jin}, {Li}, {Li}, {Li}, {Li}, {Li}, {Li}, {Li},
  {Li}, {Li}, {Li}, {Li}, {Liang}, {Liao}, {Liu}, {Liu}, {Liu}, {Liu}, {Liu},
  {Liu}, {Lu}, {Lu}, {Luo}, {Ma}, {Meng}, {Nang}, {Nie}, {Ou}, {Sai}, {Shang},
  {Sun}, {Tan}, {Tao}, {Tuo}, {Wang}, {Wang}, {Wang}, {Wang}, {Wang}, {Wen},
  {Wu}, {Wu}, {Xiao}, {Xiong}, {Xu}, {Yan}, {Yang}, {Yang}, {Yang}, {Zhang},
  {Zhang}, {Zhang}, {Zhang}, {Zhang}, {Zhang}, {Zhang}, {Zhang}, {Zhang},
  {Zhang}, {Zhang}, {Zhang}, {Zhang}, {Zhang}, {Zhang}, {Zhang}, {Zhang},
  {Zhang}, {Zhao}, {Zhao}, {Zhao}, {Zheng}, {Zhu}, {Zhu}, {Zou}, \&
  {Insight-HXMT Collaboration}}]{Huang2018}
{Huang}, Y., {Qu}, J.~L., {Zhang}, S.~N., {et~al.} 2018, \apj, 866, 122

\bibitem[{{Jithesh} {et~al.}(2021){Jithesh}, {Misra}, {Maqbool}, \&
  {Mall}}]{Jithesh2021}
{Jithesh}, V., {Misra}, R., {Maqbool}, B., \& {Mall}, G. 2021, \mnras, 505, 713

\bibitem[{{Jourdain} {et~al.}(2014){Jourdain}, {Roques}, \&
  {Chauvin}}]{Jourdain2014}
{Jourdain}, E., {Roques}, J.~P., \& {Chauvin}, M. 2014, \apj, 789, 26

\bibitem[{{Jourdain} {et~al.}(2012){Jourdain}, {Roques}, {Chauvin}, \&
  {Clark}}]{Jourdain2012}
{Jourdain}, E., {Roques}, J.~P., {Chauvin}, M., \& {Clark}, D.~J. 2012, \apj,
  761, 27

\bibitem[{{Kajava} {et~al.}(2019){Kajava}, {Motta}, {Sanna}, {Veledina}, {Del
  Santo}, \& {Segreto}}]{Kajava2019}
{Kajava}, J.~J.~E., {Motta}, S.~E., {Sanna}, A., {et~al.} 2019, \mnras, 488,
  L18

\bibitem[{{Kantzas} {et~al.}(2021){Kantzas}, {Markoff}, {Beuchert}, {Lucchini},
  {Chhotray}, {Ceccobello}, {Tetarenko}, {Miller-Jones}, {Bremer}, {Garcia},
  {Grinberg}, {Uttley}, \& {Wilms}}]{Kantzas2021}
{Kantzas}, D., {Markoff}, S., {Beuchert}, T., {et~al.} 2021, \mnras, 500, 2112

\bibitem[{{Kawamuro} {et~al.}(2018){Kawamuro}, {Negoro}, {Yoneyama}, {Ueno},
  {Tomida}, {Ishikawa}, {Sugawara}, {Isobe}, {Shimomukai}, {Mihara},
  {Sugizaki}, {Nakahira}, {Iwakiri}, {Yatabe}, {Takao}, {Matsuoka}, {Kawai},
  {Sugita}, {Yoshii}, {Tachibana}, {Harita}, {Morita}, {Yoshida}, {Sakamoto},
  {Serino}, {Kawakubo}, {Kitaoka}, {Hashimoto}, {Tsunemi}, {Nakajima},
  {Kawase}, {Sakamaki}, {Maruyama}, {Ueda}, {Hori}, {Tanimoto}, {Oda},
  {Morita}, {Yamada}, {Tsuboi}, {Nakamura}, {Sasaki}, {Kawai}, {Sato},
  {Yamauchi}, {Hanyu}, {Hidaka}, {Yamaoka}, \& {Shidatsu}}]{Kawamuro2018}
{Kawamuro}, T., {Negoro}, H., {Yoneyama}, T., {et~al.} 2018, The Astronomer's
  Telegram, 11399, 1

\bibitem[{{Labanti} {et~al.}(2002){Labanti}, {Di Cocco}, {Malaguti}, {Stephen},
  {Rossi}, {Schiavone}, {Traci}, {Ferro}, {Ferriani}, {Mauri}, \&
  {Visparelli}}]{Labanti2002}
{Labanti}, C., {Di Cocco}, G., {Malaguti}, G., {et~al.} 2002, Nuclear
  Instruments and Methods in Physics Research A, 477, 561

\bibitem[{{Lamer} {et~al.}(2020){Lamer}, {Schwope}, {Predehl}, {Traulsen},
  {Wilms}, \& {Freyberg}}]{Lamer2020}
{Lamer}, G., {Schwope}, A.~D., {Predehl}, P., {et~al.} 2020, arXiv e-prints,
  arXiv:2012.11754

\bibitem[{{Laurent} {et~al.}(2016){Laurent}, {Gouiffes}, {Rodriguez}, \&
  {Chambouleyron}}]{Laurent2016}
{Laurent}, P., {Gouiffes}, C., {Rodriguez}, J., \& {Chambouleyron}, V. 2016, in
  11th INTEGRAL Conference Gamma-Ray Astrophysics in Multi-Wavelength
  Perspective, 22

\bibitem[{{Laurent} {et~al.}(2011){Laurent}, {Rodriguez}, {Wilms}, {Cadolle
  Bel}, {Pottschmidt}, \& {Grinberg}}]{Laurent2011}
{Laurent}, P., {Rodriguez}, J., {Wilms}, J., {et~al.} 2011, Science, 332, 438

\bibitem[{{Lepingwell} {et~al.}(2018){Lepingwell}, {Bazzano}, {Bird},
  {Chenevez}, {Fiocchi}, \& {Sguera}}]{Lepingwell2018}
{Lepingwell}, V.~A., {Bazzano}, A., {Bird}, A.~J., {et~al.} 2018, The
  Astronomer's Telegram, 11884, 1

\bibitem[{{Magdziarz} \& {Zdziarski}(1995)}]{Magdziarz1995}
{Magdziarz}, P. \& {Zdziarski}, A.~A. 1995, \mnras, 273, 837

\bibitem[{{Maier} {et~al.}(2014){Maier}, {Tenzer}, \& {Santangelo}}]{Maier2014}
{Maier}, D., {Tenzer}, C., \& {Santangelo}, A. 2014, \pasp, 126, 459

\bibitem[{{Markoff} {et~al.}(2005){Markoff}, {Nowak}, \& {Wilms}}]{Markoff2005}
{Markoff}, S., {Nowak}, M.~A., \& {Wilms}, J. 2005, \apj, 635, 1203

\bibitem[{{Matsuoka} {et~al.}(2009){Matsuoka}, {Kawasaki}, {Ueno}, {Tomida},
  {Kohama}, {Suzuki}, {Adachi}, {Ishikawa}, {Mihara}, {Sugizaki}, {Isobe},
  {Nakagawa}, {Tsunemi}, {Miyata}, {Kawai}, {Kataoka}, {Morii}, {Yoshida},
  {Negoro}, {Nakajima}, {Ueda}, {Chujo}, {Yamaoka}, {Yamazaki}, {Nakahira},
  {You}, {Ishiwata}, {Miyoshi}, {Eguchi}, {Hiroi}, {Katayama}, \&
  {Ebisawa}}]{Matsuoka2009}
{Matsuoka}, M., {Kawasaki}, K., {Ueno}, S., {et~al.} 2009, \pasj, 61, 999

\bibitem[{{Miller} {et~al.}(2018){Miller}, {Gendreau}, {Ludlam}, {Fabian},
  {Altamirano}, {Arzoumanian}, {Bult}, {Cackett}, {Homan}, {Kara}, {Neilsen},
  {Remillard}, \& {Tombesi}}]{Miller2018}
{Miller}, J.~M., {Gendreau}, K., {Ludlam}, R.~M., {et~al.} 2018, \apjl, 860,
  L28

\bibitem[{{Mirabel} {et~al.}(1998){Mirabel}, {Dhawan}, {Chaty}, {Rodriguez},
  {Marti}, {Robinson}, {Swank}, \& {Geballe}}]{Mirabel1998}
{Mirabel}, I.~F., {Dhawan}, V., {Chaty}, S., {et~al.} 1998, \aap, 330, L9

\bibitem[{{Mirabel} {et~al.}(1992){Mirabel}, {Rodriguez}, {Cordier}, {Paul}, \&
  {Lebrun}}]{Mirabel1992}
{Mirabel}, I.~F., {Rodriguez}, L.~F., {Cordier}, B., {Paul}, J., \& {Lebrun},
  F. 1992, \nat, 358, 215

\bibitem[{{Mitsuda} {et~al.}(1984){Mitsuda}, {Inoue}, {Koyama}, {Makishima},
  {Matsuoka}, {Ogawara}, {Shibazaki}, {Suzuki}, {Tanaka}, \&
  {Hirano}}]{Mitsuda1984}
{Mitsuda}, K., {Inoue}, H., {Koyama}, K., {et~al.} 1984, \pasj, 36, 741

\bibitem[{{Negoro} {et~al.}(2017){Negoro}, {Ishikawa}, {Ueno}, {Tomida},
  {Sugawara}, {Isobe}, {Shimomukai}, {Mihara}, {Sugizaki}, {Serino}, {Iwakiri},
  {Shidatsu}, {Matsuoka}, {Kawai}, {Sugita}, {Yoshii}, {Tachibana}, {Harita},
  {Muraki}, {Morita}, {Yoshida}, {Sakamoto}, {Kawakubo}, {Kitaoka},
  {Hashimoto}, {Tsunemi}, {Yoneyama}, {Nakajima}, {Kawase}, {Sakamaki}, {Ueda},
  {Hori}, {Tanimoto}, {Oda}, {Tsuboi}, {Nakamura}, {Sasaki}, {Kawai},
  {Yamauchi}, {Hanyu}, {Hidaka}, {Kawamuro}, \& {Yamaoka}}]{Negoro2017}
{Negoro}, H., {Ishikawa}, M., {Ueno}, S., {et~al.} 2017, The Astronomer's
  Telegram, 10699, 1

\bibitem[{{Quinn}(2012)}]{Quinn2012}
{Quinn}, J.~L. 2012, \aap, 538, A65

\bibitem[{{Remillard} \& {McClintock}(2006)}]{Remillard2006}
{Remillard}, R.~A. \& {McClintock}, J.~E. 2006, \araa, 44, 49

\bibitem[{{Rodi} {et~al.}(2021){Rodi}, {Tramacere}, {Onori}, {Bruni},
  {S{\`a}nchez-Fern{\`a}ndez}, {Fiocchi}, {Natalucci}, \&
  {Ubertini}}]{Rodi2021}
{Rodi}, J., {Tramacere}, A., {Onori}, F., {et~al.} 2021, \apj, 910, 21

\bibitem[{{Rodriguez} {et~al.}(2015){Rodriguez}, {Grinberg}, {Laurent},
  {Cadolle Bel}, {Pottschmidt}, {Pooley}, {Bodaghee}, {Wilms}, \&
  {Gouiff{\`e}s}}]{Rodriguez2015}
{Rodriguez}, J., {Grinberg}, V., {Laurent}, P., {et~al.} 2015, \apj, 807, 17

\bibitem[{{Rodriguez} {et~al.}(2008{\natexlab{a}}){Rodriguez}, {Hannikainen},
  {Shaw}, {Pooley}, {Corbel}, {Tagger}, {Mirabel}, {Belloni}, {Cabanac},
  {Cadolle Bel}, {Chenevez}, {Kretschmar}, {Lehto}, {Paizis}, {Varni{\`e}re},
  \& {Vilhu}}]{Rodriguez2008a}
{Rodriguez}, J., {Hannikainen}, D.~C., {Shaw}, S.~E., {et~al.}
  2008{\natexlab{a}}, \apj, 675, 1436

\bibitem[{{Rodriguez} {et~al.}(2008{\natexlab{b}}){Rodriguez}, {Shaw},
  {Hannikainen}, {Belloni}, {Corbel}, {Cadolle Bel}, {Chenevez}, {Prat},
  {Kretschmar}, {Lehto}, {Mirabel}, {Paizis}, {Pooley}, {Tagger},
  {Varni{\`e}re}, {Cabanac}, \& {Vilhu}}]{Rodriguez2008b}
{Rodriguez}, J., {Shaw}, S.~E., {Hannikainen}, D.~C., {et~al.}
  2008{\natexlab{b}}, \apj, 675, 1449

\bibitem[{{Romero} {et~al.}(2014){Romero}, {Vieyro}, \& {Chaty}}]{Romero2014}
{Romero}, G.~E., {Vieyro}, F.~L., \& {Chaty}, S. 2014, \aap, 562, L7

\bibitem[{{Roques} \& {Jourdain}(2019)}]{Roques&Jourdain2019}
{Roques}, J.-P. \& {Jourdain}, E. 2019, \apj, 870, 92

\bibitem[{{Rushton} {et~al.}(2012){Rushton}, {Miller-Jones}, {Campana},
  {Evangelista}, {Paragi}, {Maccarone}, {Pooley}, {Tudose}, {Fender},
  {Spencer}, \& {Dhawan}}]{Rushton2012}
{Rushton}, A., {Miller-Jones}, J.~C.~A., {Campana}, R., {et~al.} 2012, \mnras,
  419, 3194

\bibitem[{{Russell} {et~al.}(2020){Russell}, {Lucchini}, {Tetarenko},
  {Miller-Jones}, {Sivakoff}, {Krau{\ss}}, {Mulaudzi}, {Baglio}, {Russell},
  {Altamirano}, {Ceccobello}, {Corbel}, {Degenaar}, {van den Eijnden},
  {Fender}, {Heinz}, {Koljonen}, {Maitra}, {Markoff}, {Migliari}, {Parikh},
  {Plotkin}, {Rupen}, {Sarazin}, {Soria}, \& {Wijnands}}]{Russell2020}
{Russell}, T.~D., {Lucchini}, M., {Tetarenko}, A.~J., {et~al.} 2020, \mnras,
  498, 5772

\bibitem[{{Russell} {et~al.}(2017){Russell}, {Miller-Jones}, {Sivakoff},
  {Tetarenko}, \& {Jacpot Xrb Collaboration}}]{Russell2017}
{Russell}, T.~D., {Miller-Jones}, J.~C.~A., {Sivakoff}, G.~R., {Tetarenko},
  A.~J., \& {Jacpot Xrb Collaboration}. 2017, The Astronomer's Telegram, 10711,
  1

\bibitem[{{Rybicki} \& {Lightman}(1986)}]{Rybicki&Lightman1986}
{Rybicki}, G.~B. \& {Lightman}, A.~P. 1986, {Radiative Processes in
  Astrophysics}

\bibitem[{{Shidatsu} {et~al.}(2019){Shidatsu}, {Nakahira}, {Murata}, {Adachi},
  {Kawai}, {Ueda}, \& {Negoro}}]{Shidatsu2019}
{Shidatsu}, M., {Nakahira}, S., {Murata}, K.~L., {et~al.} 2019, \apj, 874, 183

\bibitem[{{Sreehari} {et~al.}(2019){Sreehari}, {Ravishankar}, {Iyer},
  {Agrawal}, {Katoch}, {Mandal}, \& {Nandi}}]{Sreehari2019}
{Sreehari}, H., {Ravishankar}, B.~T., {Iyer}, N., {et~al.} 2019, \mnras, 487,
  928

\bibitem[{{Sridhar} {et~al.}(2019){Sridhar}, {Bhattacharyya}, {Chandra}, \&
  {Antia}}]{Sridhar2019}
{Sridhar}, N., {Bhattacharyya}, S., {Chandra}, S., \& {Antia}, H.~M. 2019,
  \mnras, 487, 4221

\bibitem[{{Stevens} {et~al.}(2018){Stevens}, {Uttley}, {Altamirano},
  {Arzoumanian}, {Bult}, {Cackett}, {Fabian}, {Gendreau}, {Ha}, {Homan},
  {Ingram}, {Kara}, {Kellogg}, {Ludlam}, {Miller}, {Neilsen}, {Pasham},
  {Remillard}, {Steiner}, \& {van den Eijnden}}]{Stevens2018}
{Stevens}, A.~L., {Uttley}, P., {Altamirano}, D., {et~al.} 2018, \apjl, 865,
  L15

\bibitem[{{Stiele} \& {Kong}(2018)}]{Stiele2018}
{Stiele}, H. \& {Kong}, A.~K.~H. 2018, \apj, 868, 71

\bibitem[{{Stirling} {et~al.}(2001){Stirling}, {Spencer}, {de la Force},
  {Garrett}, {Fender}, \& {Ogley}}]{Stirling2001}
{Stirling}, A.~M., {Spencer}, R.~E., {de la Force}, C.~J., {et~al.} 2001,
  \mnras, 327, 1273

\bibitem[{{Suffert} {et~al.}(1959){Suffert}, {Endt}, \&
  {Hoogenboom}}]{Suffert1959}
{Suffert}, M., {Endt}, P.~M., \& {Hoogenboom}, A.~M. 1959, Physica, 25, 659

\bibitem[{{Tao} {et~al.}(2018){Tao}, {Chen}, {G{\"u}ng{\"o}r}, {Huang}, {Lu},
  {Qu}, {Song}, {Zhang}, {Zhang}, \& {Zhang}}]{Tao2018}
{Tao}, L., {Chen}, Y., {G{\"u}ng{\"o}r}, C., {et~al.} 2018, \mnras, 480, 4443

\bibitem[{{Tetarenko} {et~al.}(2021){Tetarenko}, {Casella}, {Miller-Jones},
  {Sivakoff}, {Paice}, {Vincentelli}, {Maccarone}, {Gandhi}, {Dhillon},
  {Marsh}, {Russell}, \& {Uttley}}]{Tetarenko2021}
{Tetarenko}, A.~J., {Casella}, P., {Miller-Jones}, J.~C.~A., {et~al.} 2021,
  \mnras, 504, 3862

\bibitem[{{Tominaga} {et~al.}(2020){Tominaga}, {Nakahira}, {Shidatsu}, {Oeda},
  {Ebisawa}, {Sugawara}, {Negoro}, {Kawai}, {Sugizaki}, {Ueda}, \&
  {Mihara}}]{Tominaga2020}
{Tominaga}, M., {Nakahira}, S., {Shidatsu}, M., {et~al.} 2020, \apjl, 899, L20

\bibitem[{{Torres} {et~al.}(2020){Torres}, {Casares}, {Jim{\'e}nez-Ibarra},
  {{\'A}lvarez-Hern{\'a}ndez}, {Mu{\~n}oz-Darias}, {Armas Padilla}, {Jonker},
  \& {Heida}}]{Torres2020}
{Torres}, M.~A.~P., {Casares}, J., {Jim{\'e}nez-Ibarra}, F., {et~al.} 2020,
  \apjl, 893, L37

\bibitem[{{Torres} {et~al.}(2019){Torres}, {Casares}, {Jim{\'e}nez-Ibarra},
  {Mu{\~n}oz-Darias}, {Armas Padilla}, {Jonker}, \& {Heida}}]{Torres2019}
{Torres}, M.~A.~P., {Casares}, J., {Jim{\'e}nez-Ibarra}, F., {et~al.} 2019,
  \apjl, 882, L21

\bibitem[{{Trushkin} {et~al.}(2018){Trushkin}, {Nizhelskij}, {Tsybulev}, \&
  {Erkenov}}]{Trushkin2018}
{Trushkin}, S.~A., {Nizhelskij}, N.~A., {Tsybulev}, P.~G., \& {Erkenov}, A.
  2018, The Astronomer's Telegram, 11539, 1

\bibitem[{{Ubertini} {et~al.}(2003){Ubertini}, {Lebrun}, {Di Cocco}, {Bazzano},
  {Bird}, {Broenstad}, {Goldwurm}, {La Rosa}, {Labanti}, {Laurent}, {Mirabel},
  {Quadrini}, {Ramsey}, {Reglero}, {Sabau}, {Sacco}, {Staubert}, {Vigroux},
  {Weisskopf}, \& {Zdziarski}}]{Ubertini2003}
{Ubertini}, P., {Lebrun}, F., {Di Cocco}, G., {et~al.} 2003, \aap, 411, L131

\bibitem[{{Vaillancourt}(2006)}]{Vaillancourt2006}
{Vaillancourt}, J.~E. 2006, \pasp, 118, 1340

\bibitem[{{Vedrenne} {et~al.}(2003){Vedrenne}, {Roques}, {Sch{\"o}nfelder},
  {Mand rou}, {Lichti}, {von Kienlin}, {Cordier}, {Schanne}, {Kn{\"o}dlseder},
  {Skinner}, {Jean}, {Sanchez}, {Caraveo}, {Teegarden}, {von Ballmoos},
  {Bouchet}, {Paul}, {Matteson}, {Boggs}, {Wunderer}, {Leleux},
  {Weidenspointner}, {Durouchoux}, {Diehl}, {Strong}, {Cass{\'e}}, {Clair}, \&
  {Andr{\'e}}}]{Vedrenne2003}
{Vedrenne}, G., {Roques}, J.~P., {Sch{\"o}nfelder}, V., {et~al.} 2003, \aap,
  411, L63

\bibitem[{{Vincentelli} {et~al.}(2021){Vincentelli}, {Casella}, {Russell},
  {Baglio}, {Veledina}, {Maccarone}, {Malzac}, {Fender}, {O'Brien}, \&
  {Uttley}}]{Vincentelli2021}
{Vincentelli}, F.~M., {Casella}, P., {Russell}, D.~M., {et~al.} 2021, \mnras,
  503, 614

\bibitem[{{Wang} {et~al.}(2020){Wang}, {Ji}, {Zhang}, {M{\'e}ndez}, {Qu},
  {Maggi}, {Ge}, {Qiao}, {Tao}, {Zhang}, {Altamirano}, {Zhang}, {Ma}, {Lu},
  {Li}, {Huang}, {Zheng}, {Chen}, {Chang}, {Tuo}, {G{\"u}ng{\"o}r}, {Song},
  {Xu}, {Cao}, {Chen}, {Liu}, {Bu}, {Cai}, {Chen}, {Chen}, {Chen}, {Chen},
  {Cui}, {Cui}, {Deng}, {Dong}, {Du}, {Fu}, {Gao}, {Gao}, {Gao}, {Gu}, {Guan},
  {Guo}, {Han}, {Huo}, {Jia}, {Jiang}, {Jiang}, {Jin}, {Jin}, {Kong}, {Li},
  {Li}, {Li}, {Li}, {Li}, {Li}, {Li}, {Li}, {Li}, {Li}, {Liang}, {Liao}, {Liu},
  {Liu}, {Liu}, {Liu}, {Lu}, {Lu}, {Luo}, {Luo}, {Meng}, {Nang}, {Nie}, {Ou},
  {Sai}, {Shang}, {Song}, {Sun}, {Tan}, {Wang}, {Wang}, {Wang}, {Wang}, {Wang},
  {Wen}, {Wu}, {Wu}, {Wu}, {Xiao}, {Xiao}, {Xiong}, {Yang}, {Yang}, {Yang},
  {Yang}, {Yi}, {Yin}, {You}, {Zhang}, {Zhang}, {Zhang}, {Zhang}, {Zhang},
  {Zhang}, {Zhang}, {Zhang}, {Zhang}, {Zhang}, {Zhang}, {Zhang}, {Zhang},
  {Zhang}, {Zhang}, {Zhang}, {Zhao}, {Zhao}, {Zhou}, {Zhou}, {Zhuang}, {Zhu},
  {Zhu}, \& {Wang}}]{Wang2020}
{Wang}, Y., {Ji}, L., {Zhang}, S.~N., {et~al.} 2020, \apj, 896, 33

\bibitem[{{Weisskopf} {et~al.}(2006){Weisskopf}, {Elsner}, {Hanna}, {Kaspi},
  {O'Dell}, {Pavlov}, \& {Ramsey}}]{Weisskopf2006}
{Weisskopf}, M.~C., {Elsner}, R.~F., {Hanna}, D., {et~al.} 2006, arXiv
  e-prints, astro

\bibitem[{{Wilms} {et~al.}(2000){Wilms}, {Allen}, \& {McCray}}]{Wilms2000}
{Wilms}, J., {Allen}, A., \& {McCray}, R. 2000, \apj, 542, 914

\bibitem[{{Xu} {et~al.}(2018){Xu}, {Harrison}, {Garc{\'\i}a}, {Fabian},
  {F{\"u}rst}, {Gandhi}, {Grefenstette}, {Madsen}, {Miller}, {Parker},
  {Tomsick}, \& {Walton}}]{Xu2018}
{Xu}, Y., {Harrison}, F.~A., {Garc{\'\i}a}, J.~A., {et~al.} 2018, \apjl, 852,
  L34

\bibitem[{{Yatabe} {et~al.}(2019){Yatabe}, {Negoro}, {Nakajima}, {Sakamaki},
  {Maruyama}, {Aoki}, {Kobayashi}, {Mihara}, {Nakahira}, {Takao}, {Matsuoka},
  {Sakamoto}, {Serino}, {Sugita}, {Hashimoto}, {Yoshida}, {Kawai}, {Sugizaki},
  {Tachibana}, {Morita}, {Ueno}, {Tomida}, {Ishikawa}, {Sugawara}, {Isobe},
  {Shimomukai}, {Midooka}, {Ueda}, {Tanimoto}, {Morita}, {Yamada}, {Ogawa},
  {Tsuboi}, {Iwakiri}, {Sasaki}, {Kawai}, {Sato}, {Tsunemi}, {Yoneyama},
  {Asakura}, {Ide}, {Yamauchi}, {Hidaka}, {Iwahori}, {Kawamuro}, {Yamaoka},
  {Shidatsu}, \& {Kawakubo}}]{Yatabe2019}
{Yatabe}, F., {Negoro}, H., {Nakajima}, M., {et~al.} 2019, The Astronomer's
  Telegram, 12425, 1

\bibitem[{{Zdziarski} {et~al.}(1996){Zdziarski}, {Johnson}, \&
  {Magdziarz}}]{Zdziarski1996}
{Zdziarski}, A.~A., {Johnson}, W.~N., \& {Magdziarz}, P. 1996, \mnras, 283, 193

\bibitem[{{Zdziarski} {et~al.}(2021){Zdziarski}, {Jourdain}, {Lubi{\'n}ski},
  {Szanecki}, {Nied{\'z}wiecki}, {Veledina}, {Poutanen}, {Dzie{\l}ak}, \&
  {Roques}}]{Zdziarski2021}
{Zdziarski}, A.~A., {Jourdain}, E., {Lubi{\'n}ski}, P., {et~al.} 2021, \apjl,
  914, L5

\bibitem[{{Zdziarski} {et~al.}(2020){Zdziarski}, {Shapopi}, \&
  {Pooley}}]{Zdziarski2020}
{Zdziarski}, A.~A., {Shapopi}, J.~N.~S., \& {Pooley}, G.~G. 2020, \apjl, 894,
  L18

\bibitem[{{Zhang} {et~al.}(2021){Zhang}, {Altamirano}, {Uttley}, {Garc{\'\i}a},
  {M{\'e}ndez}, {Homan}, {Steiner}, {Alabarta}, {Buisson}, {Remillard},
  {Gendreau}, {Arzoumanian}, {Markwardt}, {Strohmayer}, {Neilsen}, \&
  {Basak}}]{Zhang2021}
{Zhang}, L., {Altamirano}, D., {Uttley}, P., {et~al.} 2021, \mnras, 505, 3823

\end{thebibliography}

\newpage

\begin{appendix}

\section{Additional figures}

\begin{figure*}[h]
    \centering
    \includegraphics[width=\columnwidth]{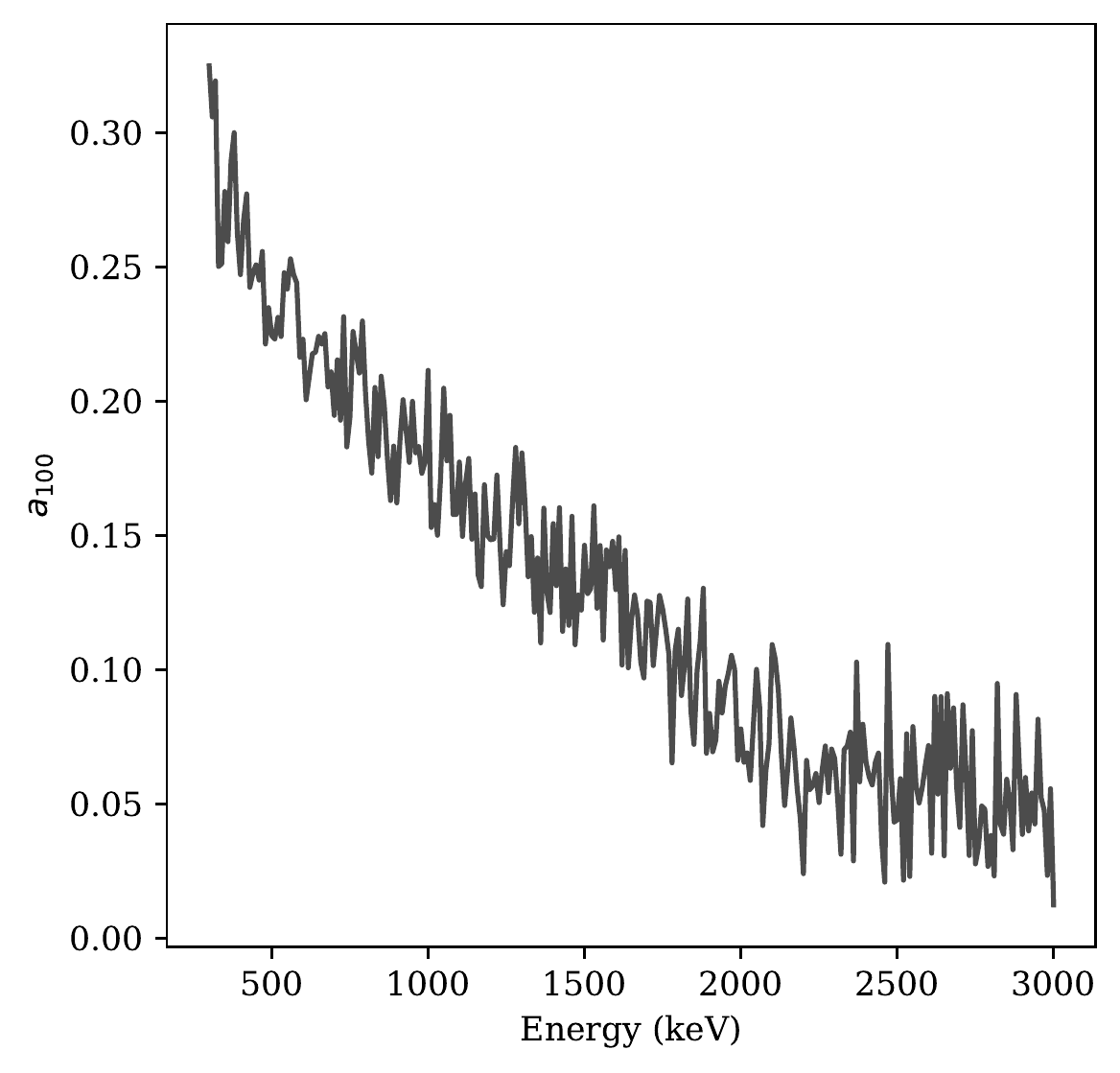}
    \caption{Evolution of the modulation parameter $a_\mathrm{100}$ as a function of the energy.}
    \label{fig:modulation}
\end{figure*}

\begin{figure*}[h]
    \centering
    \includegraphics[width=\textwidth]{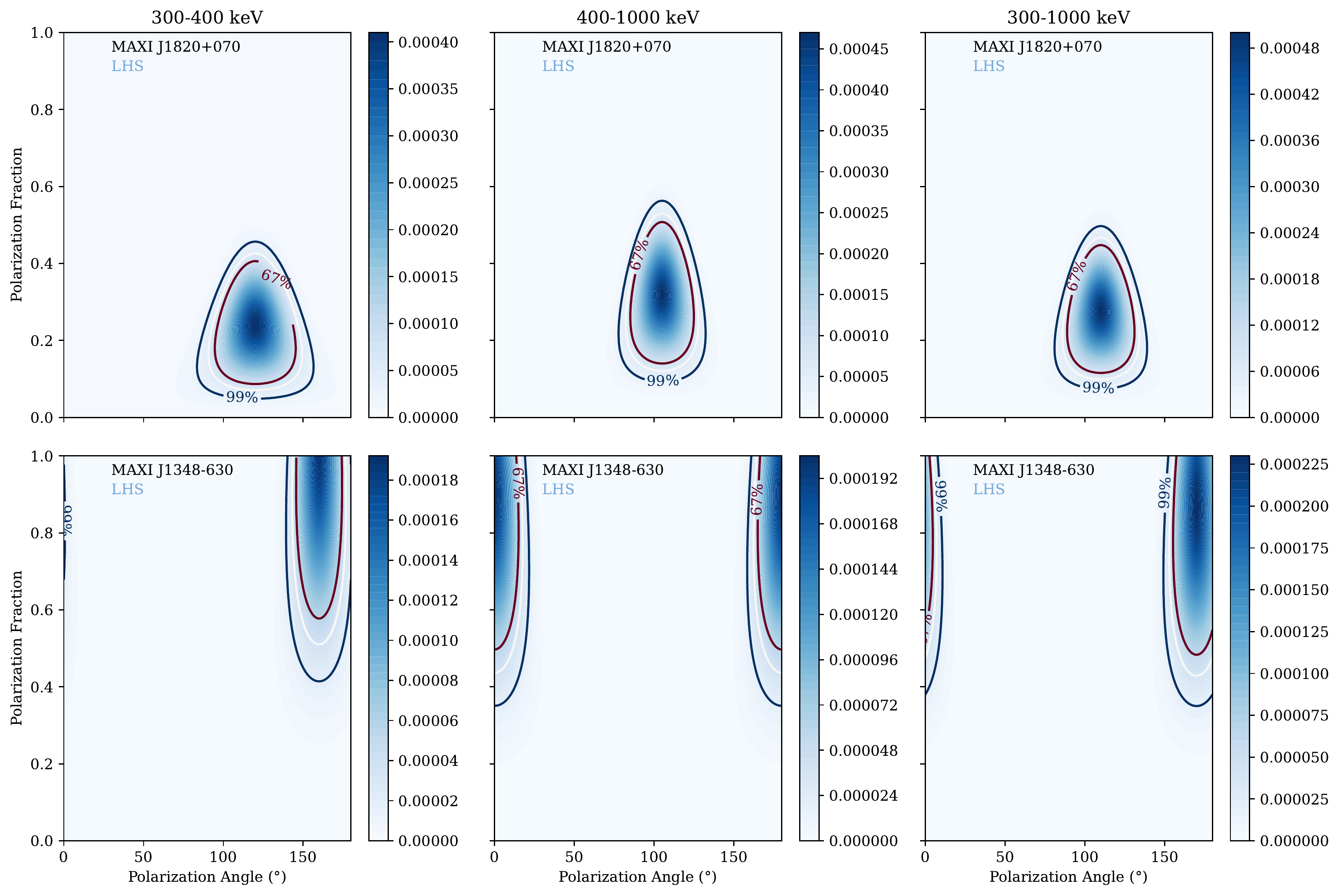}
    \caption{Probability density described by equation \eqref{eq:proba_pola} in function of the polarization angle and the polarization fraction calculated for \dixhuit (top) and \treize (bottom) in the LHS for different energy bands.}
    \label{fig:all_contours}
\end{figure*}

\begin{figure*}
    \centering
    \includegraphics[width=0.94\textwidth]{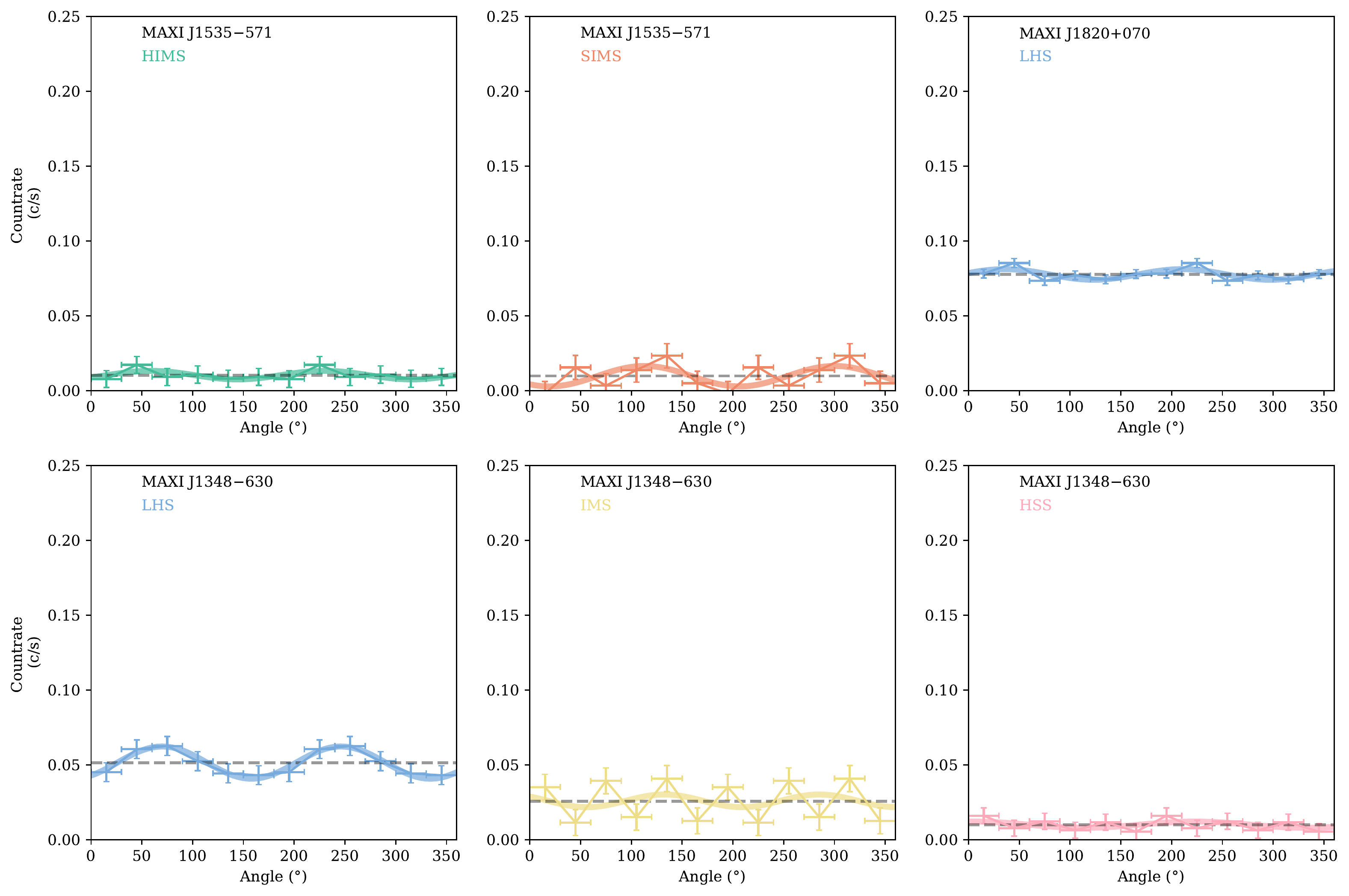}
    \caption{Polarigrams of the three sources obtained in the 300--400\,keV energy band range.}
    \label{fig:pola_300-400}
\end{figure*}

\begin{figure*}
    \centering
    \includegraphics[width=0.94\textwidth]{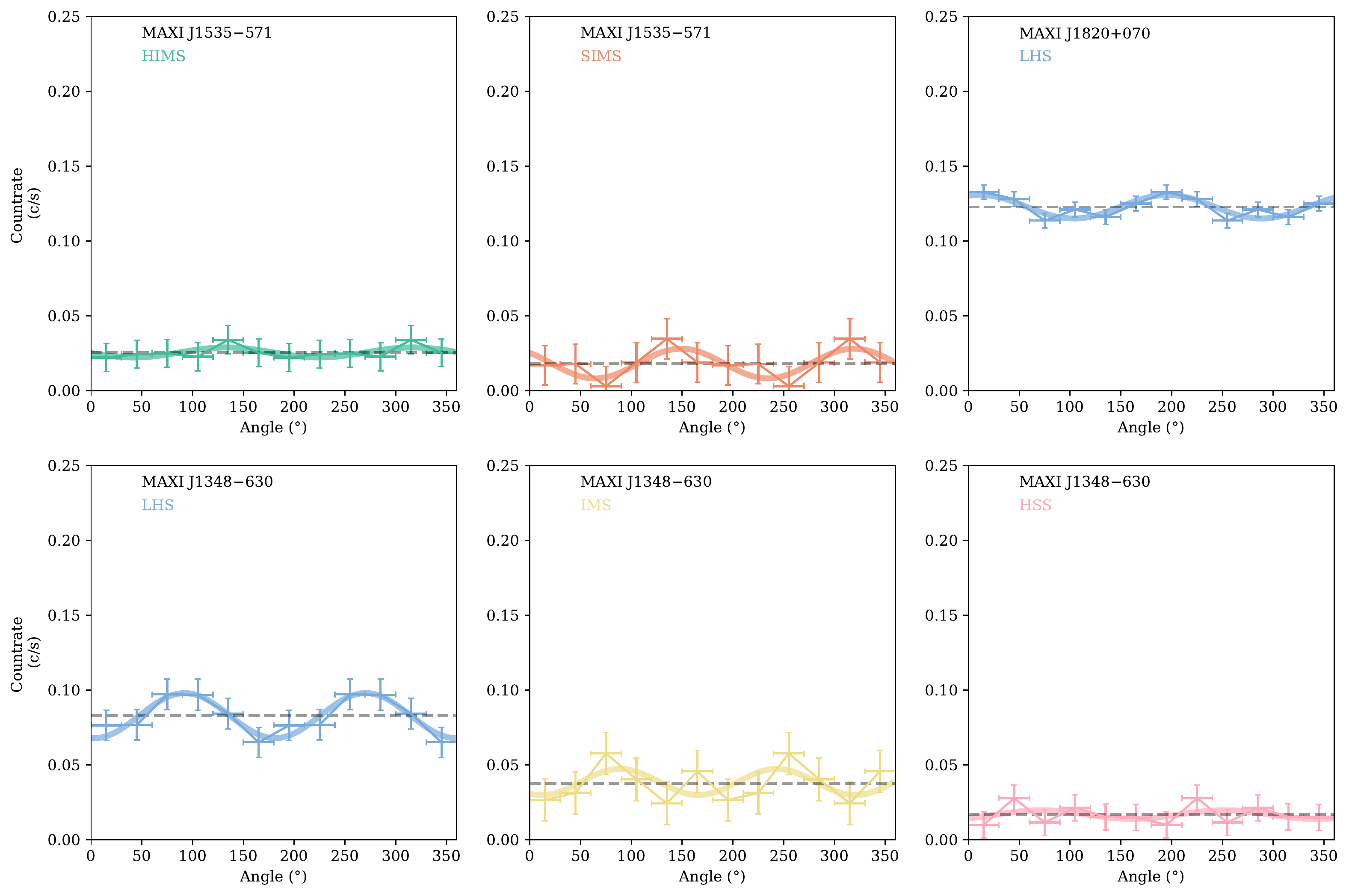}
    \caption{Polarigrams of the three sources obtained in the 400--1000\,keV energy band range.}
    \label{fig:pola_400-1000}
\end{figure*}


\end{appendix}
 
¨

\end{document}